%% file: main.tex
\documentclass{article}

\usepackage{arxiv}

\usepackage[utf8]{inputenc} 
\usepackage[T1]{fontenc}    
\usepackage{hyperref}       
\usepackage{url}            
\usepackage{booktabs}       
\usepackage{amsfonts}       
\usepackage{nicefrac}       
\usepackage{microtype}      
\usepackage{amsmath} 
\usepackage{amssymb} 
\usepackage{cleveref}       
\usepackage{graphicx}
\usepackage{natbib}
\usepackage{doi}
\usepackage{subcaption} 
\usepackage{multirow}   
\usepackage{algorithm, algpseudocode}

\title{Learning Normal Patterns in Musical Loops}

\date{}

\newif\ifuniqueAffiliation
\uniqueAffiliationtrue

\ifuniqueAffiliation
\author{ \href{https://orcid.org/0000-0003-1970-5353}{\includegraphics[scale=0.06]{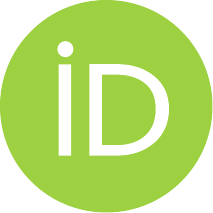}\hspace{1mm}Shayan Dadman}\thanks{Corresponding author.} \\
	Department of Computer Science\\
	UiT, The Arctic University of Tromsø\\
	Lodve Langesgate 2, 8514 Narvik, Norway \\
	\texttt{shayan.dadman@uit.no} \\
	\And
	Bernt Arild Bremdal \\
	Department of Computer Science\\
	UiT, The Arctic University of Tromsø\\
	Lodve Langesgate 2, 8514 Narvik, Norway \\
	\texttt{bernt.a.bremdal@uit.no} \\
	\AND
	Børre Bang \\
	Department of Computer Science\\
	UiT, The Arctic University of Tromsø\\
	Lodve Langesgate 2, 8514 Narvik, Norway \\
	\texttt{borre.bang@uit.no} \\
	\And
	\href{https://orcid.org/0000-0001-6070-303X}{\includegraphics[scale=0.06]{orcid.pdf}\hspace{1mm}Rune Dalmo} \\
	Department of Computer Science\\
	UiT, The Arctic University of Tromsø\\
	Lodve Langesgate 2, 8514 Narvik, Norway \\
	\texttt{rune.dalmo@uit.no} \\
}
\else

\usepackage{authblk}

\newbox{\orcid}\sbox{\orcid}{\includegraphics[scale=0.06]{orcid.pdf}} 
\author[1]{%
	\href{https://orcid.org/0000-0003-1970-5353}{\usebox{\orcid}\hspace{1mm}Shayan Dadman\thanks{\texttt{shayan.dadman@uit.no}}}%
}
\author[1]{%
	Bernt Arild Bremdal\thanks{\texttt{bernt.a.bremdal@uit.no}}%
}
\author[1]{%
	\href{https://orcid.org/0000-0000-0000-0000}{\usebox{\orcid}\hspace{1mm}Børre Bang\thanks{\texttt{borre.bang@uit.no}}}%
}
\author[1]{%
	\href{https://orcid.org/0000-0001-6070-303X}{\usebox{\orcid}\hspace{1mm}Rune Dalmo\thanks{\texttt{rune.dalmo@uit.no}}}%
}
\affil[1]{Department of Computer Science, UiT, The Arctic University of Tromsø}
\fi

\renewcommand{\undertitle}{}

\hypersetup{
pdftitle={Learning Normal Patterns in Musical Loops: Unsupervised Anomaly Detection for Variable-Length Audio Inputs},
pdfsubject={cs.AI, cs.LG},
pdfauthor={Shayan Dadman, Bernt Arild Bremdal, Børre Bang, Rune Dalmo},
pdfkeywords={First keyword, Second keyword, More},
}

\begin{document}
\maketitle

\begin{abstract}
	This paper introduces an unsupervised framework for detecting audio patterns in musical samples (loops) through anomaly detection techniques, addressing challenges in music information retrieval (MIR). Existing methods are often constrained by reliance on handcrafted features, domain-specific limitations, or dependence on iterative user interaction. We address these limitations through an architecture combining deep feature extraction with unsupervised anomaly detection. Our approach leverages a pre-trained Hierarchical Token-semantic Audio Transformer (HTS-AT), paired with a Feature Fusion Mechanism (FFM), to generate representations from variable-length audio loops. These embeddings are processed using one-class Deep Support Vector Data Description (Deep SVDD), which learns normative audio patterns by mapping them to a compact latent hypersphere. Evaluations on curated bass and guitar datasets compare standard and residual autoencoder variants against baselines like Isolation Forest (IF) and and principle component analysis (PCA) methods. Results show our Deep SVDD models, especially the residual autoencoder variant, deliver improved anomaly separation, particularly for larger variations. This research contributes a flexible, fully unsupervised solution for processing diverse audio samples, overcoming previous structural and input limitations while enabling effective pattern identification through distance-based latent space scoring.
\end{abstract}


\section{Introduction}\label{sec:introduction}

Musical loops are building blocks in modern music production, especially within genres like Hip-Hop and Electronic music \citep{collins_electronic_2014, butler_unlocking_2006}. These seamlessly repeatable audio segments provide essential rhythmic, melodic, or harmonic structures \citep{reese_audio_2009}. Music creators, including producers and DJs, often navigate vast libraries or extract segments from existing tracks to find or create loops that fit their artistic vision \citep{gibson_smrt_2005, streich2008music}. Efficiently analyzing patterns within these loops is essential, not only for selecting compatible elements \citep{kitahara2015loop, chen_neural_2020} but also for identifying variations, distinct sonic characteristics, or potential inconsistencies within extensive collections or generated sequences.

Beyond mere efficiency, the evolving role of Artificial Intelligence (AI) in music necessitates tools that empower creators through interactivity, adaptability, and user control. As highlighted in broader discussions on AI-driven music generation systems (MGS) \citep{deruty2022Development, huang2020AI}, there is a demand for technologies that move beyond "black-box" paradigms \citep{dadman2022Interactive}. This demand is directly aligned with the central objective of our research: transforming MGS into adaptive, interactive creative partners. Our earlier work \citep{dadman2024Crafting, dadman2023multiAgent} began to address this by architecting an MGS framework envisioned for co-creation, where AI agents and human creator could collaboratively navigate the creative process.

Building upon that foundational pursuit of human-AI partnership, this paper delves into a component for such systems: the nuanced understanding and analysis of musical loops (segments). Enhancing the system's capability to engage with these building blocks of contemporary music is important for it to act as a truly insightful and supportive partner. Music creators, in particular, seek systems that can be tailored to their artistic visions and workflows, operate effectively with their private data collections, and offer transparency in their processes, as emphasized by \citet{dadman2022Interactive, civit2022Systematic}. The unsupervised pattern detection approach explored in this paper aligns with this trajectory. We hypothesize that by incorporating this approach, we can further develop its capacity to adapt to individual user styles and support a more fluid, creatively rich workflow, thereby contributing to the core hypothesis regarding AI's potential as a creative collaborator. This approach also provides opportunities for integration into current music creation workflows to facilitate creative exploration (more on this in Section \ref{sec:applications}).

Within Music Information Retrieval (MIR) \citep{fu2011Survey, ras2010Advances}, music loop detection can be viewed as a specialized audio pattern recognition task \citep{kong_panns_2020}. Audio pattern recognition, encompassing sub-tasks like audio classification and tagging \citep{xie_zero-shot_2020, gemmeke_audio_2017, choi_automatic_2016}, aims to automatically identify and categorize audio signals. This framing allows us to leverage advancements in broader audio analysis, particularly those driven by deep learning. These methods typically extract meaningful features from audio to train models for recognizing patterns or irregularities \citep{dashsukantakumar2023Comprehensive}. Notably, the success of pre-trained audio encoders in generalizing across diverse datasets \citep{wu_large-scale_2024} underpins our hypothesis that such models can form the basis for loop analysis, particularly when designed to capture nuanced temporal and hierarchical information.

Despite advances in audio pattern recognition, particularly driven by deep learning \citep{kong_panns_2020, wu_large-scale_2024}, existing approaches for audio loop analysis often face limitations. Traditional methods relying on signal processing and music structure analysis techniques \citep{lu_repeating_2004, ong2006structural, paulus2010state, nieto_audio-based_2020} can struggle with the complexity and variability of real-world music, often requiring heuristic tuning. More recent methods, especially those employing neural networks (NNs), impose constraints such as requiring fixed-length inputs \citep{chen_neural_2020, han_symbolic_2022}. This limits their applicability to the inherently variable durations of musical loops encountered in real-world scenarios. Furthermore, some systems necessitate iterative user feedback \citep{jakubik_searching_2022} or extensive labeled data, hindering their automation potential and scalability. Consequently, there remains a need for methods that can analyze variable-length audio loops, adapt to diverse musical structures, and operate effectively without extensive supervision or domain-specific constraints.

To address these gaps and advance the development of more adaptable AI tools, this paper proposes an unsupervised method that frames audio loop pattern detection as an anomaly detection task. The core insight is that 'typical' or 'normative' loop patterns, characteristic of a given dataset or style, can be learned implicitly from unlabeled data. This unsupervised characteristic is key, as it allows the system to adapt to individual user collections and specific stylistic nuances without requiring extensive pre-labeled datasets, thereby facilitating more personalized analyses. Deviations from these learned norms can then be identified as anomalies, providing creators with insights—highlighting not only variations or potential errors for automated quality control but also distinct stylistic elements that could spark creative discovery and serendipitous sound exploration (discussed further in Section \ref{sec:discussion}).

Our methodology implements this unsupervised anomaly detection framework by integrating a pre-trained audio feature extractor with an anomaly detection module. Specifically, we employ a Hierarchical Token-semantic Audio Transformer (HTS-AT) \citep{chen_hts-at_2022}, chosen for its demonstrated ability to capture both local and global temporal dependencies in audio. To handle variable-length inputs—a challenge identified earlier—the HTS-AT is augmented with a Feature Fusion Mechanism (FFM) \citep{wu_large-scale_2024}, which together, they generate fixed-dimension embeddings that encapsulate temporal and hierarchical audio information. A Deep Support Vector Data Description (Deep SVDD) network \citep{ruff_deep_2018} is then trained on these embeddings. Deep SVDD was selected for its capability to learn a compact hypersphere that encloses 'normal' data instances in a feature space. We demonstrate that by learning this compact hypersphere of 'normal' loop patterns from the feature representations, Deep SVDD can effectively identify loops outside this boundary, thereby providing a reliable and adaptable solution for loop pattern detection and analysis.

The main contributions of this work are:

\begin{itemize}
  \item We introduce and implement an unsupervised framework for audio loop pattern detection, combining HTS-AT with FFM for flexible feature extraction, and Deep SVDD for anomaly-based identification directly from the learned audio embeddings. This approach offers a pathway to data-driven insights from user-specific collections, aligning with the need for more adaptable AI tools.
  \item We demonstrate the system's capability to process variable-length audio inputs through the FFM, thereby overcoming a limitation of many prior fixed-length approaches and enhancing practical applicability.
  \item We provide an empirical evaluation on curated datasets of bass and guitar loops, showcasing the model's proficiency in learning representations of normative patterns and effectively identifying meaningful deviations.
  \item We conduct a comparative analysis of different Deep SVDD encoder architectures (a standard Autoencoder versus an Autoencoder with Residual Connections), offering insights into architectural choices that benefit the modeling of diverse and larger audio data.
  \item We benchmark our proposed method against standard unsupervised anomaly detection techniques (Isolation Forest (IF) and PCA-based reconstruction error), demonstrating the enhanced representational power and discriminative ability of the Deep SVDD approach when operating on HTS-AT embeddings.
\end{itemize}

The remainder of this paper is structured as follows: Section \ref{sec:related_works} reviews relevant prior work. Section \ref{sec:methodology} details the proposed architecture and its components. Section \ref{sec:experimental_setup} describes the datasets, preprocessing steps, and training procedures. Section \ref{sec:evaluation} outlines the evaluation and baseline methods. Section \ref{sec:results} presents the experimental results. Section \ref{sec:discussion} discusses the findings and their implications, followed by Section \ref{sec:limitations_future_work}, which addresses limitations and outlines future research directions. Finally, Section \ref{sec:conclusion} concludes this paper.

\section{Related Works}\label{sec:related_works}

\subsection{Audio Pattern Recognition}
Advancements in deep learning have impacted audio pattern recognition tasks. Large-scale Pretrained Audio Neural Networks (PANNs) \citep{kong_panns_2020}, trained on extensive datasets like AudioSet \citep{gemmeke_audio_2017}, serve as powerful feature extractors applicable to various downstream tasks, including audio tagging, acoustic scene classification, and sound event detection \citep{xu2023SemiSupervised}. Models like Convolutional Neural Networks (CNNs), Residual Networks (ResNets) \citep{he_deep_2016}, and MobileNets \citep{howard2017MobileNets} have been adapted for these tasks, often using analysis in frequency domain (Mel-spectrograms) as input features \citep{kong_panns_2020}. While some approaches explore 1D CNNs operating directly on waveforms (e.g., DaiNet, LeeNet), they have generally not surpassed spectrogram-based methods, potentially due to difficulties in capturing frequency-related patterns \citep{kong_panns_2020}. \citeauthor{kong_panns_2020} proposed the Wavegram-Logmel-CNN, combining learned time-frequency representations (Wavegrams) with log-mel spectrograms to achieve state-of-the-art results on AudioSet tagging. Their work demonstrated the benefit of integrating learned and hand-crafted features. Furthermore, techniques like FFM have been explored to address challenges such as variable-length input limitations in audio processing models \citep{wu_large-scale_2024}, showing efficacy across multiple datasets and suggesting potential applicability in loop detection. Architectures like the HTS-AT \citep{wu_large-scale_2024} represent recent developments in modeling sequential audio data by offering improved performance in pattern recognition tasks, compared to PANNs.

\subsection{Loop Selection and Extraction}
Research targeting music loops often focuses on identifying or extracting compatible loops from existing tracks. These methods historically employed various signal processing and rule-based techniques. Early approaches for loop extraction often relied on finding autocorrelation peaks in handcrafted harmonic or spectral features (e.g., chroma, Mel-frequency Cepstral Coefficients - MFCCs) to detect repeating patterns \citep{shi2018LoopMaker, ong2008Music}. Other methods estimated loop similarity using features representing harmony, timbre, and energy, as noted by \citeauthor{chen_neural_2020}. For assessing loop compatibility, rule-based systems like AutoMashUpper \citep{davies2014AutoMashUpper} computed similarity based on features like chromagrams, rhythm patterns, and spectral balance, while other techniques explored psychoacoustic principles or tonal interval spaces \citep{gebhardt2016Psychoacoustic, bernardes2017Perceptuallymotivateda}. Decomposition methods, such as Nonnegative Tensor Factorization (NTF), aimed to separate loops directly from audio mixes by modeling tracks as combinations of repeating templates and activation patterns \citep{smith2018Nonnegative}. While useful, these methods often depend on the choice of handcrafted features, may require heuristic tuning, and can struggle with the complexity and variability of real-world music \citep{chen_neural_2020}. Motivated by the limitations of these traditional approaches, particularly in capturing complex compatibility relationships and generalizing across diverse audio, recent research has increasingly focused on leveraging NNs.

\citet{chen_neural_2020} addressed the challenge of finding compatible loops from large libraries by proposing NN models to estimate compatibility. They investigated a CNN approach, which evaluates the combined time-frequency representation of a loop pair, and a Siamese Neural Network (SNN) architecture, which processes loops separately before comparing embeddings. Their models were trained on loop pairs extracted from Hip-Hop tracks within the Free Music Archive (FMA) dataset \cite{defferrard_fma_2016}. A preprocessing step involved time-stretching all loops to a fixed duration (2 seconds) and representing them as log mel-spectrograms before feeding them into the networks \citep{chen_neural_2020}. While their learned models outperformed a rule-based baseline (AutoMashUpper \citep{davies2014AutoMashUpper}) in subjective tests \citep{chen_neural_2020}, the reliance on fixed-length inputs limits flexibility in handling loops of varying durations commonly found in practice.

\citet{jakubik_searching_2022} focused on retrieving subjectively interesting loops and sound samples within electronic music tracks using an active learning system. This approach relies on user interaction to progressively refine search results, starting from a single user-provided example \citep{jakubik_searching_2022}. The study evaluated feature learning, comparing standard MFCCs against features learned unsupervisedly via methods combining autoencoders and contrastive learning (specifically Bootstrap Your Own Latent - BYOL \citep{grill_bootstrap_2020}). The dataset comprised electronic music tracks from \texttt{sampleswap}\footnote{\url{https://www.sampleswap.org/}}, annotated for loops and samples. While feature learning improved recall compared to MFCCs, particularly when trained on representative data, the system's performance varied across genres (e.g., struggling with Dubstep) and fundamentally depends on iterative user feedback, limiting fully automatic application \citep{jakubik_searching_2022}. Furthermore, according to authors, finding distinct sound samples proved harder than finding repeating loops.

\citet{han_symbolic_2022} explored symbolic music loop generation, focusing on creating 8-bar bass and drum loops in MIDI format. They proposed a two-stage process: first compressing MIDI loop data into discrete latent codes using a Vector Quantized Variational Autoencoder (VQ-VAE), and then training an autoregressive model on these codes to generate new loops \citep{han_symbolic_2022}. At the core of their approach is a cross-domain loop detector. Their loop detector was trained on 1,000 audio loops from \texttt{looperman}\footnote{\url{https://www.looperman.com/}} using One-Class Deep SVDD \citep{ruff_deep_2018}, learning to identify domain-invariant structural patterns in 8-bar phrases. They transformed the audio loops into bar-to-bar (8-bar) correlation matrices to train the loop detector to identify appropriate 8-bar segments. This detector was then applied to extract MIDI loops from the Lakh MIDI Dataset\footnote{\url{https://colinraffel.com/projects/lmd/}}. However, this method has limitations similar to \cite{chen_neural_2020}. The fixed 8-bar window restricts the system's flexibility to accommodate phrases of varying lengths, and the structural analysis primarily on bar-to-bar correlations may struggle to capture higher-level musical nuances present in audio recordings.

\subsection{Comparative Analysis and Motivation}

The reviewed literature highlights persistent challenges in automated loop analysis and detection: processing efficiency for real-time use, flexibility in handling variable-length audio inputs without information loss or requiring fixed-size representations, and adaptability to diverse musical structures beyond fixed patterns or specific domains \citep{han_symbolic_2022, jakubik_searching_2022, chen_neural_2020}. Our proposed methodology aims to address these gaps by leveraging a dual-component architecture consisting of an Audio Encoder with FFM and HTS-AT model, and a Deep SVDD module for anomaly detection to directly process encoded signals. This allows for detection and analysis of loops with diverse characteristics directly from audio and alleviate the structural rigidity and constraints of prior methods.

\section{Methodology}\label{sec:methodology}

\begin{figure}[t]
\centering
\includegraphics[width=\textwidth]{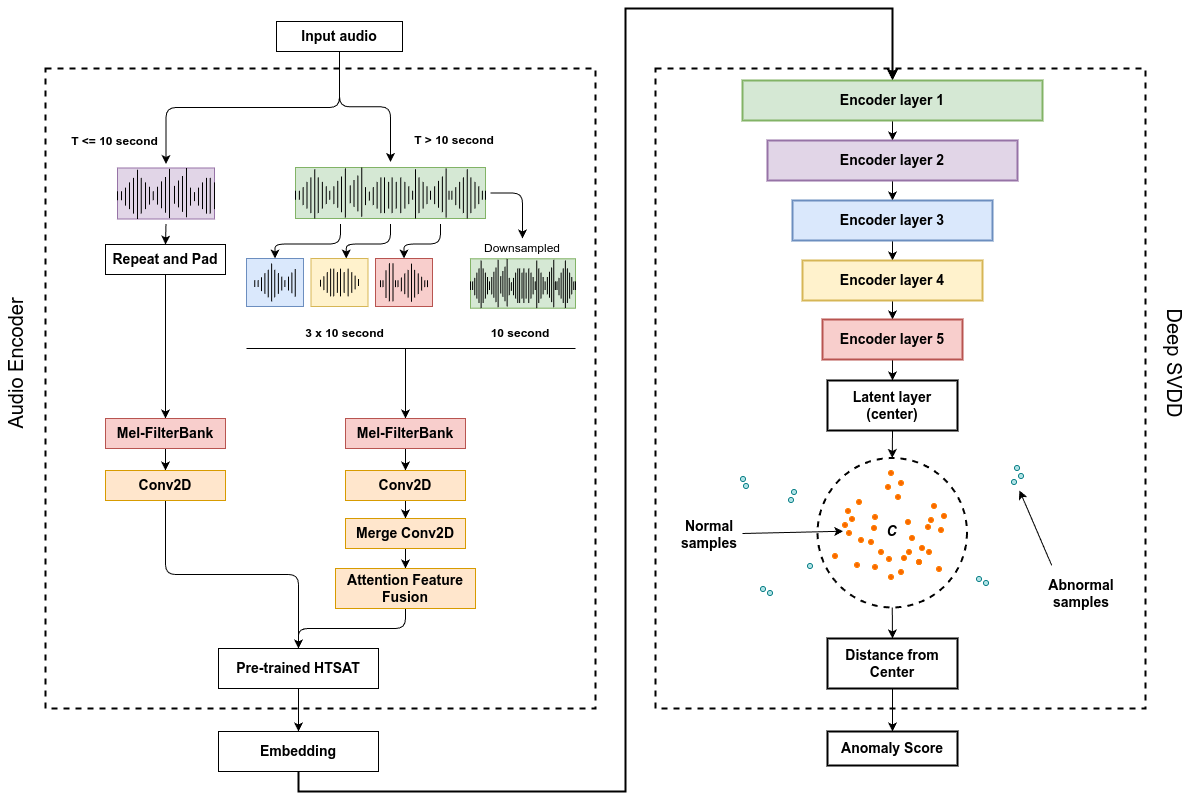}
\caption{Architectural overview of the proposed loop detection model. The model comprises two main components: (left) an Audio Encoder module for hierarchical feature extraction from audio inputs and (right) a Deep SVDD module for anomaly detection. The Audio Encoder processes input audio through dual pathways based on duration, utilizing Mel-FilterBank processing, Conv2D layers, the feature fusion mechanism, and a pre-trained HTS-AT for feature extraction. The Deep SVDD module consists of five encoder layers that transform features into a latent space, where normal samples (orange dots) are mapped within a hypersphere and abnormal samples (blue dots) are mapped outside. The architecture enables end-to-end training for unsupervised loop detection and anomaly identification through distance-based scoring from the learned latent representation. The illustration of the Audio Encoder is inspired by \citet{wu_large-scale_2024}. \label{fig:model_architecture}}
\end{figure}

This section details the proposed methodology for identifying and characterizing repetitive patterns within audio loops. Our core strategy reframes this challenge as an anomaly detection task: We aim to learn a representation of 'normal' loop structures so that significant deviations or less common patterns can be effectively identified as 'anomalies'. As depicted in Figure \ref{fig:model_architecture}, our system architecture comprises two sequential stages: an Audio Encoder and a Deep SVDD module. The audio encoder generates a compact embedding from the input audio by leveraging the FFM and HTS-AT pre-trained model. The Deep SVDD module then processes this embedding.

The Audio Encoder employs a dual-path processing strategy dictated by the input audio duration, designed to handle variable input lengths, while capturing both fine-grained local details and coarse global structures. For inputs of 10 seconds or less, a repeat and pad mechanism ensures consistent input dimensions. For longer inputs, the audio is divided into three 10-second segments (start, middle, end) for local analysis, alongside a globally downsampled 10-second version for broader context. Regardless of the path, initial processing involves converting the audio waveform into Mel-spectrograms using a Mel-Filter Bank. These spectral features then pass through initial Convolutional 2D (Conv2D) layers. For longer inputs, features from the three local segments are merged through a dedicated Conv2D layer and subsequently combined with the global features using an Attention Feature Fusion (AFF) mechanism (detailed in Section \ref{sec:feature_fusion}). The resulting features, representing either the padded short input or the fused multi-scale representation of the long input, are fed into the HTS-AT model (Section \ref{sec:hts_at}). The output of the Audio Encoder serves as the final audio embedding passed to the Deep SVDD stage.

The embeddings generated by the Audio Encoder are then processed by the Deep SVDD module (Section \ref{sec:deep_svdd}), comprising five sequential encoder layers. These layers map the high-dimensional embedding from the Audio Encoder into a lower-dimensional latent space. The objective within this space is to learn a hypersphere that encloses representations of 'normal' audio patterns while representations of 'anomalous' or distinct patterns fall outside this boundary. The distance of an embedding from the hypersphere's center provides a score for pattern identification or anomaly detection. The following subsections detail the implementation and role of each key component within this architecture.

\subsection{Feature Fusion Mechanism}\label{sec:feature_fusion}
To address the challenge of variable audio input lengths, a limitation noted in prior works (Section \ref{sec:related_works}), we implement the FFM inspired by \citeauthor{wu_large-scale_2024}. This mechanism specifically handles the processing divergence between various audio clip lengths within the Audio Encoder. It ensures consistent processing dimensions and integrates information from different temporal scales. For audio clips of fixed chunk duration \textit{d}-seconds or less, the input Mel-spectrograms are repeated and padded to meet the \textit{d}-second requirement. For clips longer than \textit{d}-seconds, a dual representation is created:

\begin{enumerate}
  \item Global: The entire clip is downsampled to a \textit{d}-second representation.
  \item Local: Three \textit{d}-second segments are sampled (start, middle, end).
\end{enumerate}

Initial features are extracted from the Mel-spectrograms of these representations using Conv2D layers. For the long clips, features from the three local segments are then consolidated using an additional Conv2D layer.

The core of the fusion for long clips is the Attention Feature Fusion (AFF) module, adapted from \citeauthor{dai_attentional_2020}. The AFF takes the extracted global features and the consolidated local features as input. It dynamically computes attention scores based on multi-scale contextual analysis of these features, employing point-wise convolutions and non-linear activations. These scores determine a fusion ratio, adaptively weighting the contribution of global versus local information to generate a single, fused feature map. For a detailed explanation of the fusion mechanism's internal workings, please refer to \citet{wu_large-scale_2024}. The resulting fused map, compromising both broad structure and specific details, serves as the input to the subsequent HTS-AT module.

\subsection{Hierarchical Token-semantic Audio Transformer}\label{sec:hts_at}
We selected the HTS-AT, proposed by \citeauthor{chen_hts-at_2022}, as the core feature extractor within our Audio Encoder. This choice is motivated by HTS-AT's demonstrated efficiency and efficacy in modeling complex, hierarchical structures inherent in audio data, offering potential advantages over standard transformers or PANNs in audio classification tasks \citet{xin2024Audiotext, xin2023Backgroundaware, xin2023Improving}.

In our implementation, the HTS-AT model inputs the feature maps generated by the preceding stages (either from padded short clips or the FFM for long clips). The model processes these features through its hierarchical transformer blocks, which utilize windowed attention to manage computational complexity while effectively capturing temporal dependencies at multiple scales. A key aspect of HTS-AT is its patch-merging mechanism between blocks, which progressively reduces sequence length, further enhancing efficiency. Unlike models relying solely on a classification (CLS) token, HTS-AT's token-semantic module provides finer-grained outputs. However, this work primarily utilizes the final aggregated representation as the audio embedding.

For this study, we utilize the pre-trained HTS-AT model weights provided at \texttt{HuggingFace} \footnote{\url{https://huggingface.co/lukewys/laion_clap/tree/main}}, trained on a combination of music, Audioset, and LAION-Audio-630k datasets. The weights are kept frozen during training of the Deep SVDD module. The output of the HTS-AT model is the final, fixed-dimension audio embedding that encapsulates the hierarchical and temporal characteristics of the input audio. For a detailed explanation of the HTS-AT's internal workings, please refer to \citet{chen_hts-at_2022}.

\subsection{Deep Support Vector Data Description}\label{sec:deep_svdd}
The final stage of our proposed system leverages Deep SVDD \citep{ruff_deep_2018}, an unsupervised anomaly detection algorithm derived from the classical SVDD \citep{tax_support_2004}. We employ Deep SVDD for unsupervised audio loop pattern analysis by framing it as an anomaly detection problem: common, frequently occurring, or structurally coherent loop patterns within a given collection are considered 'normal,' while distinct, novel, or significantly divergent patterns are identified as 'anomalies.'

The choice of Deep SVDD is driven by several advantages for our task. Firstly, its unsupervised nature allows the model to learn directly from vast and unlabeled collections of audio loops. This aligns with our goal of creating systems adaptable to individual user libraries and diverse musical styles without requiring laborious manual annotation. Secondly, Deep SVDD learns a compact, data-driven boundary of normality. By training a neural network, specifically five sequential encoder layers as outlined in Figure \ref{fig:model_architecture}, we aim to map the fixed-dimension audio embeddings from the Audio Encoder module into a minimal-volume hypersphere. This process allows the network to effectively learn the shared characteristics of the 'normal' loop data. Such adaptability makes Deep SVDD suitable for capturing normative structures in varied musical data, a capability supported by its application in related musical context by \citet{han_symbolic_2022}.

The Deep SVDD module takes the fixed-dimension audio embeddings $\mathbf{z}$ (output by Audio Encoder module) as input. The network, denoted by $\phi(\cdot; W)$ with parameters $W$, maps these embeddings into a latent output space. The training objective, following \citeauthor{ruff_deep_2018}, is to minimize the volume of a hypersphere defined by a center $c$ and radius $R > 0$, while ensuring that the representations $\phi(\mathbf{z}; W)$ of most 'normal' patterns are enclosed within it. This can be formulated as:

\begin{equation}
\min_{W} \frac{1}{N}\sum_{i=1}^{N} \left\| \phi(z_i; W) - c \right\|^2 + \frac{\lambda}{2} \sum_{l=1}^{L} \left\| W_l \right\|_F^2
\end{equation}

where the first term penalizes distances from the center $c$ for $N$ normal training samples, and the second term is a network weight decay regularizer (with $L$ layers and Frobenius norm $|\cdot|_F$). The center $c$ is often fixed as the mean of initial network outputs for the training data or can be learned.

During training on representative 'normal' loop patterns, the network $\phi$ is optimized to learn the common factors of variation, effectively pulling their latent representations towards the hypersphere's center $c$. Consequently, loop patterns that deviate significantly from these learned commonalities will be mapped further from $c$ in the latent space. The anomaly score $S(\mathbf{x})$ for a given input loop $\mathbf{x}$ (which yields embedding $\mathbf{z}$) is then its squared Euclidean distance to the center $c$: 

\begin{equation}
S(x)=\|\phi(\mathbf{x})-c\|^2
\end{equation}

A lower score indicates that the loop's characteristics closely resemble the 'normal' patterns learned during training, while a higher score signifies a deviation, marking it as 'anomalous' or distinct. This score is the final output of the Deep SVDD module, providing a quantifiable measure to identify potentially interesting, unique, or structurally divergent audio loops within a collection.

\section{Experimental Setup}\label{sec:experimental_setup}

\subsection{Data}\label{sec:dataset}
The dataset for this study was curated from audio samples provided by MusicRadar \citep{musicradar_sampleradar}, a resource offering diverse, royalty-free audio samples in WAV format across various genres. The samples are free to use, however, redistribution restrictions prevent sharing the curated dataset directly.

Our curation focused on guitar and bass samples for simplicity, resulting in a dataset comprising 1816 bass samples (approx. 2.34 hours) and 4294 guitar samples (approx. 5.92 hours), totaling 6110 samples. The initial selection process involved downloading relevant sample packs and manually filtering them. This manual selection was performed to ensure samples possessed clear rhythmic structure, minimal background noise, and were representative of typical guitar/bass patterns. 

\begin{table}[t]
  \centering
  \caption{Statistics per instrument of the curated dataset}
  \label{tab:stats_per_instrument}
  \begin{tabular}{llcc}
  \toprule
    & Instrument & Bass & Guitar \\
  \midrule 
    & Count & 1816 & 4294 \\
    & Durations (hrs) & 2.337 & 5.924 \\
  \midrule
  \multirow{8}{*}{BPM}
    & mean & 111.779 & 108.785 \\
    & std & 18.854 & 18.000 \\
    & min & 64.000 & 73.000 \\
    & 25\% & 95.000 & 95.000 \\
    & 50\% & 114.000 & 110.000 \\
    & 75\% & 123.000 & 120.000 \\
    & max & 170.000 & 170.000 \\
  \midrule
  \multirow{8}{*}{Duration (secs)}
    & mean & 4.633 & 4.967 \\
    & std & 2.436 & 1.971 \\
    & min & 0.546 & 0.417 \\
    & 25\% & 3.692 & 4.000 \\
    & 50\% & 4.364 & 4.571 \\
    & 75\% & 5.333 & 5.647 \\
    & max & 28.346 & 22.700 \\
  \bottomrule
  \end{tabular}
\end{table}

Metadata, including genre, key mode, and beats per minute (BPM), were extracted to better characterize the dataset. Folder and file names provided initial cues, particularly for genre and key\footnote{The sample files within the sample packs provided by MusicRadar often are placed in directories with meaningful and representative names, which facilitated in extracting the corresponding metadata.}. BPM values were refined using the tempo estimator algorithm by \citet{foroughmand2019deeprhythm}, utilizing the \texttt{deeprhythm} library\footnote{\url{https://github.com/bleugreen/deeprhythm}}; the estimated tempo was adopted if it considerably (\textgreater 10) differed from the extracted metadata. Subsequently, sample durations were calculated using \texttt{librosa}\footnote{\url{https://librosa.org/doc/latest/index.html}} library.

\begin{figure}[t]
	\centering
	\begin{subfigure}[b]{0.49\linewidth}
		\includegraphics[width=\linewidth]{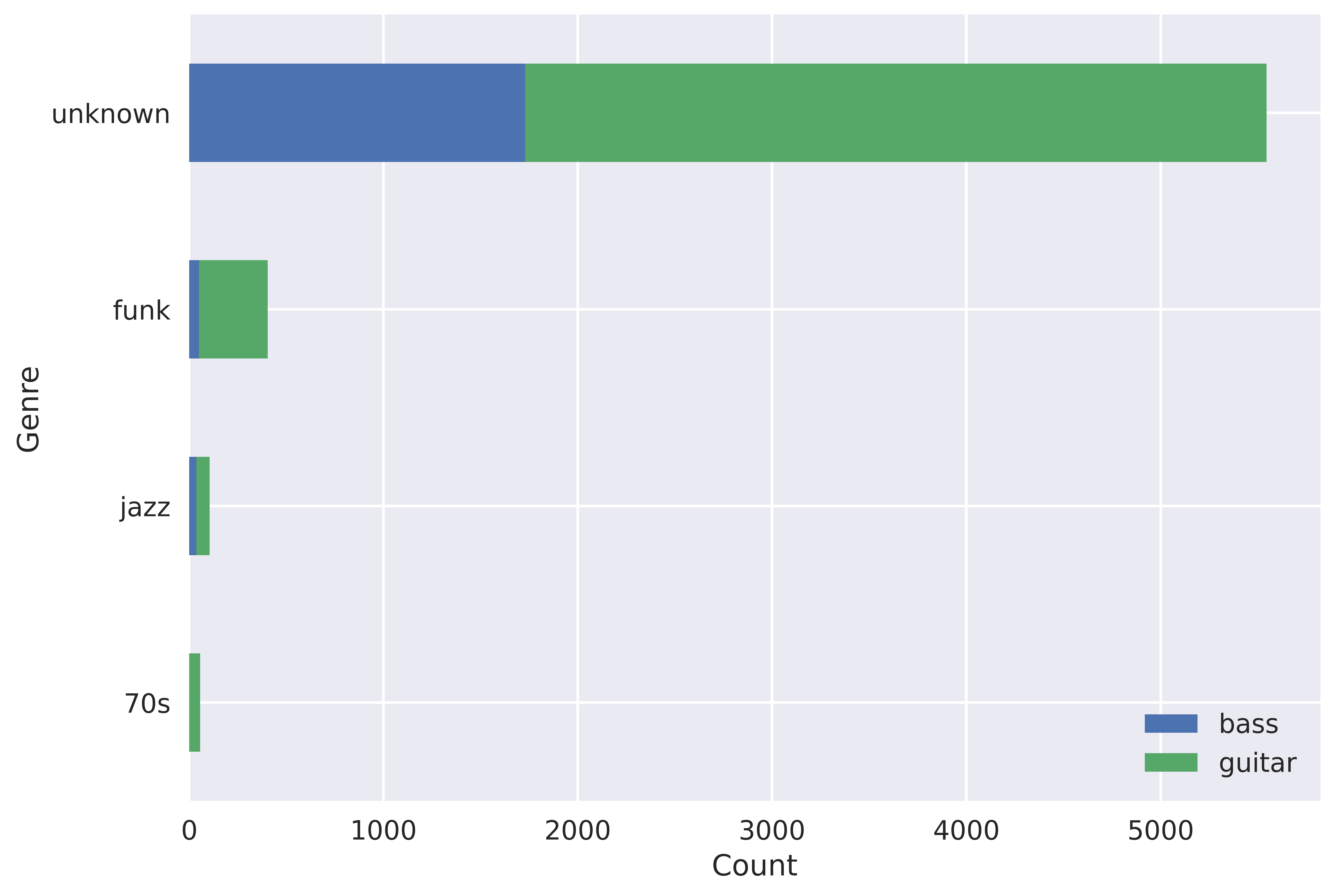}
		\caption{}
	\end{subfigure}
	\hfill
	\begin{subfigure}[b]{0.49\linewidth}
		\includegraphics[width=\linewidth]{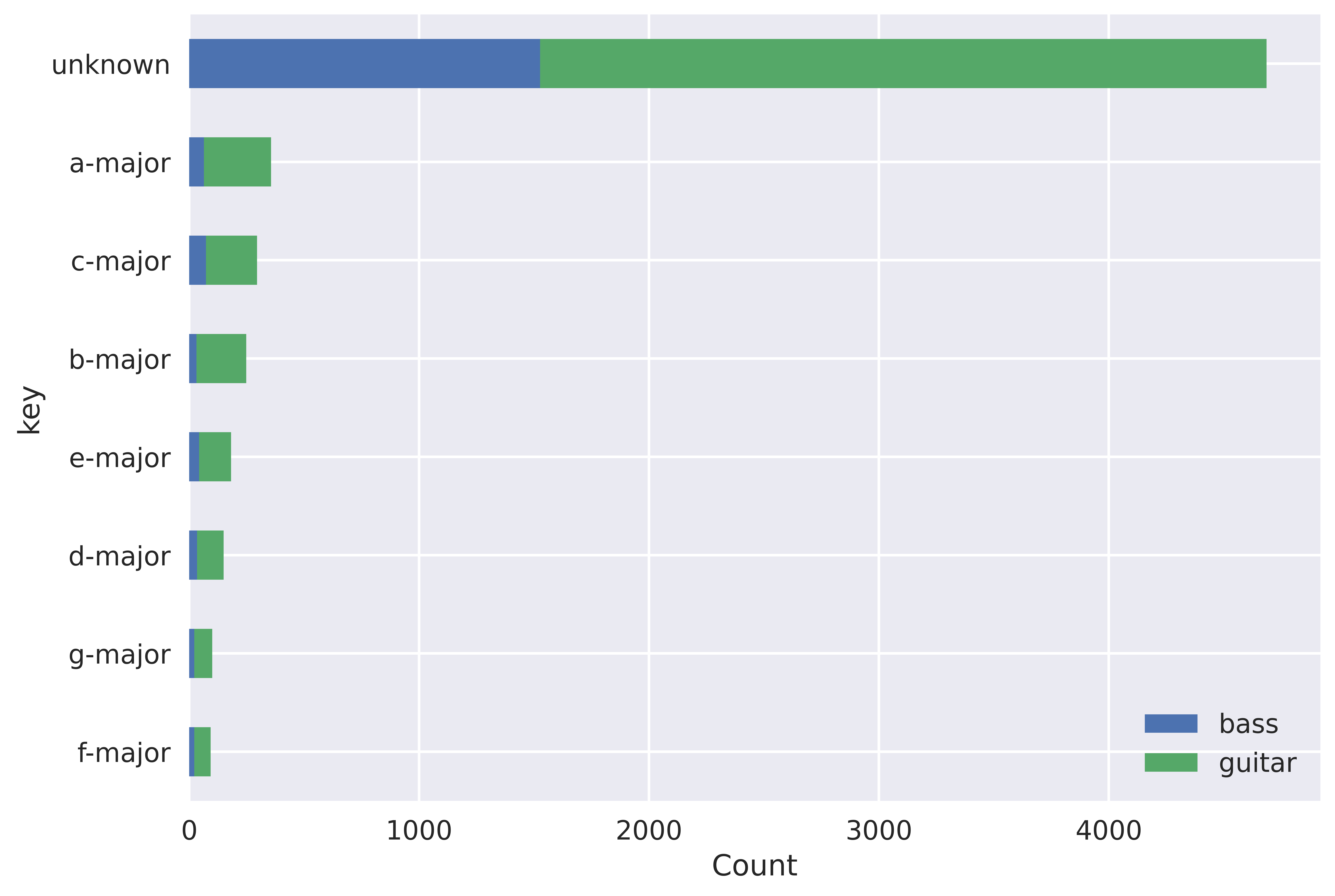}
		\caption{}
	\end{subfigure}
	\caption{Distribution of curated guitar and bass samples across (a) musical genres and (b) musical keys. The height of the bars indicates the count of samples for each category, with colors differentiating between bass (blue) and guitar (green). The prevalence of the 'unknown' category highlights common challenges with sample library metadata.}
	\label{fig:meta_genre_key}
  \end{figure}

Table \ref{tab:stats_per_instrument} provides statistics on tempo and duration per instrument, while Figure \ref{fig:meta_genre_key} illustrates the distribution across genres and musical keys, broken down by instrument. As depicted in Figure \ref{fig:meta_genre_key}a, the dataset includes samples labeled with specific genres such as Funk, Jazz, and '70s', but the vast majority of both guitar and bass samples fall into an 'unknown' genre category, reflecting common inconsistencies or omissions in sample packs metadata. Guitar samples constitute the larger portion within most categories, consistent with the overall dataset composition. Similarly, Figure \ref{fig:meta_genre_key}b illustrates the distribution by key. While several major keys (e.g., A-major, C-major, B-major) are represented in smaller numbers, the predominant category for both instruments is again 'unknown'. This distribution highlights the real-world challenge of dealing with potentially incomplete metadata in audio sample collections. Despite the prevalence of 'unknown' tags, this dataset, consisting of variations of standard guitar and bass loops, provides a suitable testbed for evaluating the model's ability to learn representations distinguishing typical, repeating patterns (treated as 'normal') from potentially distinct or 'anomalous' ones based on the audio content itself.

We considered two data subsets for our experiments: one containing only bass samples and one containing only guitar samples. In each case, we applied an 80/20 split for training and validation. The validation set was used exclusively for hyperparameter tuning and early stopping during the Deep SVDD training phases. No separate final test set was reserved at this study stage, as our primary focus was on exploring unsupervised model behavior and representation learning.

We experimented with each instrument separately, primarily due to their distinct characteristics. Bass samples typically feature lower frequencies, richer harmonic content, and different rhythmic patterns than guitar samples. In contrast, guitar samples often exhibit more mid-to-high frequency information and a broader array of tonal textures. These differences can influence the performance of machine learning models. Additionally, analyzing each instrument individually enabled us to better understand how each feature set affects Deep SVDD training. It is also important to note that even within each instrument category, there might be nuanced differences in the sound characteristics. These subtleties, which are presented in Section \ref{sec:results} and will be discussed in Section \ref{sec:discussion}, further justify the need for separate experimental analyses.

\subsection{Data Preprocessing}
The data preprocessing pipeline prepares the audio samples for input into the model architecture described in Section \ref{sec:methodology}. All audio samples were first resampled to 48 kHz. Following the methodology outlined by \cite{wu_large-scale_2024} and adapted in our approach, Mel-spectrograms were computed using a Short-Time Fourier Transform (STFT) with a window size of 1024, a hop size of 480, and 64 Mel filter banks. This configuration results in spectral representations corresponding to $(T=1024, F=64)$ dimensions for a standard 10-second audio segment. 

Input handling strictly followed the dual-path strategy defined in Section \ref{sec:feature_fusion} to accommodate variable audio lengths. Finally, the pre-trained HTS-AT audio encoder (Section \ref{sec:hts_at}) was utilized in \textbf{inference mode only}. Its weights were kept frozen throughout all experiments. The output embeddings from this frozen encoder, having a dimensionality of $(1, 1024)$, were pre-computed for all samples in the training and validation sets. These embeddings served as the direct input to the Deep SVDD module.

\subsection{Hyperparameters and Training Details}

Here, we describe the NN architectures employed in our Deep SVDD implementation, focusing on two variants: a standard Autoencoder (\textit{AE}) and an Autoencoder with Residual Connections (\textit{AEwRES}). The base \textit{AE} architecture consists of multiple fully connected layers utilizing the ELU (Exponential Linear Unit) activation function \cite{rasamoelina_review_2020}. Each layer is complemented by dropout for regularization and batch normalization to stabilize training and accelerate convergence.

\begin{figure}[t]
\centering
\includegraphics[width=\textwidth]{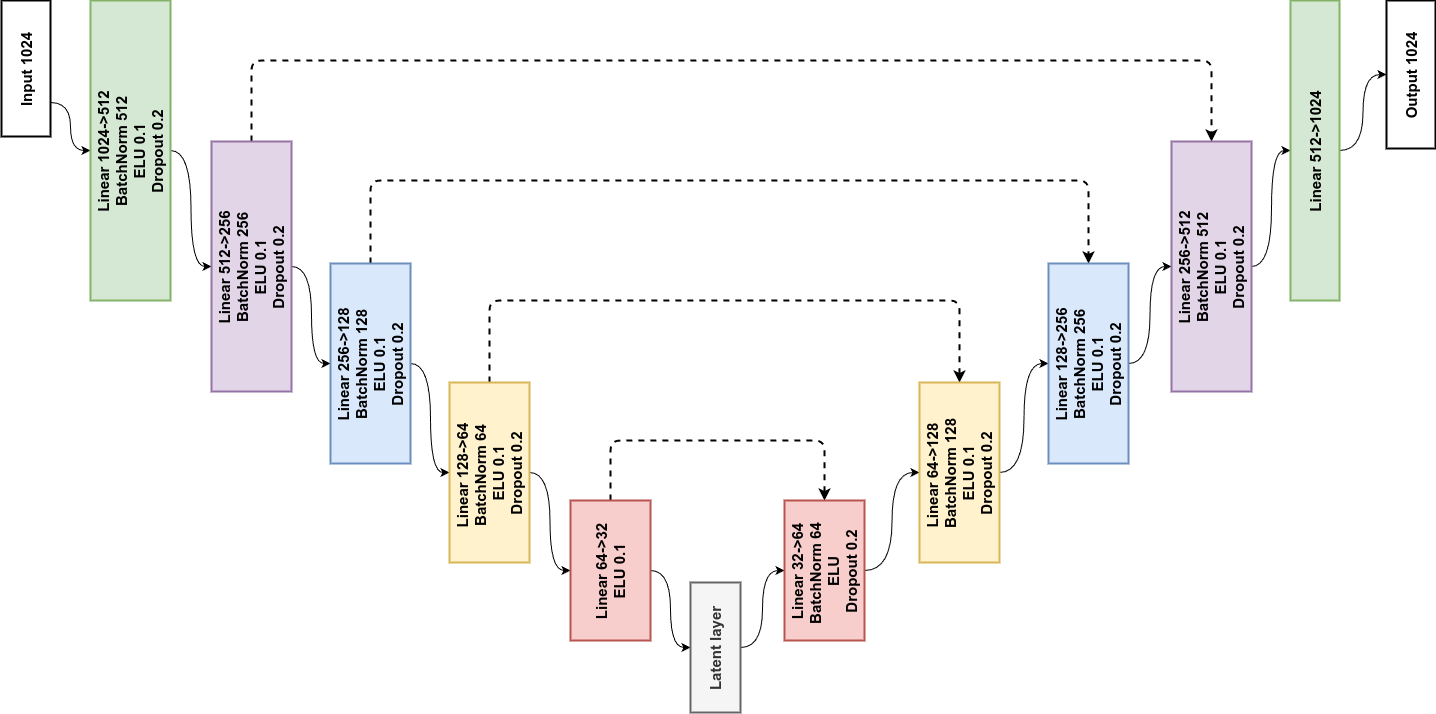}
\caption{The architecture of the autoencoder (\textit{AE}) network with residual connections (dotted arrows) (\textit{AEwRES}). The model follows a symmetric design (inspired by U-net architecture \citep{ronneberger2015UNet}) with an encoder (left) and decoder (right). The input dimension of 1024 is progressively reduced through five encoder layers and then reconstructed through five decoder layers. Each layer comprises a linear transformation followed by BatchNorm, ELU activation (\textit{alpha}=0.1), and Dropout (rate=0.2). Residual connections add the output of encoder layer $i$ to the input of decoder layer $5-i$. Note that the base \textit{AE} architecture is identical to that of \textit{AEwRES}, but without any residual connections. \label{fig:aewres_architecture}}
\end{figure}

The training procedure for the experiments followed a two-phase approach: 

\begin{enumerate} 
\item AE Pre-training: The autoencoder (either \textit{AE} or \textit{AEwRES}) was first trained on the pre-computed embeddings of the training set samples using a standard Mean Squared Error (MSE) reconstruction loss. This phase encourages the encoder to learn a compressed representation, capturing the principal variations within the data. 
\item Deep SVDD Fine-tuning: After pre-training, the decoder part was discarded. The pre-trained encoder weights were retained and fine-tuned using the Deep SVDD objective \citep{ruff_deep_2018}. This phase minimizes the volume of a hypersphere in the latent space intended to enclose the embeddings of the 'normal' training data, using a loss based on the distance of embeddings from the hypersphere center $c$. 
\end{enumerate}

For both training phases, we employed the AdamW optimizer \citep{loshchilov2019Decoupled} with an initial learning rate of $1 \times 10^{-3}$ and weight decay of $1 \times 10^{-5}$. A Cosine Annealing learning rate scheduler \citep{loshchilov2017SGDR} was used to gradually decrease the learning rate to a minimum of $5 \times 10^{-6}$ throughout training. The models were trained with a batch size of 32 for 1000 epochs. Early stopping was implemented with patience of 20 epochs, monitoring the validation loss (reconstruction loss during pre-training, sphere volume/center distance loss during fine-tuning) to prevent overfitting and select the best-performing model checkpoint based on the validation set. The experiments were implemented using the \texttt{PyTorch} framework, and experiment tracking, including hyperparameter logging and monitoring of training/validation metrics, was managed using \texttt{Weights and Biases}\footnote{\url{https://wandb.ai/site/}}.

\section{Evaluation}\label{sec:evaluation}

Evaluating anomaly detection without ground-truth labels necessitates focusing on representation quality and model behavior. We assess our proposed Deep SVDD models (standard Autoencoder - \textit{AE}, and with Residual Connections - \textit{AEwRES}), both using frozen HTS-AT embeddings. Our evaluation centers on analyzing their learned latent spaces and output anomaly scores to determine which architecture is more effective at distinguishing between normal and anomalous audio loops.

\subsection{Baseline Models}\label{sec:baseline_models}

To contextualize the performance of our proposed Deep SVDD approach, we benchmark our models against two baseline methods: IF and PCA-based reconstruction error. These baselines were selected to provide a comparative context, contrasting our approach with a general-purpose anomaly detection algorithm and a simpler, feature-based reconstruction method. This allows us to understand the benefits and drawbacks of our method relative to existing techniques. Isolation Forest \citep{liu2008isolation} serves as a general-purpose anomaly detection baseline. By comparing against this tree-based ensemble method, we can assess whether the deep feature learning in our approach offers a tangible advantage over established, non-parametric techniques. This comparison highlights the importance of learned representations for loop detection and analysis tasks.

PCA-based reconstruction error \citep{jolliffe2002principal} is employed to evaluate the benefits of our non-linear feature learning over a simpler linear manifold modeling approach. By training PCA on normal data and using the reconstruction error of new instances as an anomaly score, we determine whether anomalies are adequately captured by deviations from the primary linear correlations within the dataset. This comparison specifically highlights the added value of the non-linear transformations learned by our models and the more refined decision boundary enabled by the Deep SVDD objective. This contrasts our integrated framework with a straightforward linear approach to identifying irregularities (PCA reconstruction error).

The baselines use the same embeddings by Audio Encoder module as our models. Using the subsequent analysis techniques, this ensures a fair comparison and enables a clearer understanding of our method's specific contributions.

\subsection{PCA Projection Visualization}\label{sec:pca_projection}

To visualize the high-dimensional latent spaces in 2D, we apply PCA. For the proposed Deep SVDD models, PCA is applied to their final latent representations; for baselines, it is applied to the inputs provided by the Audio Encoder module. We generate scatter plots using the first two principal components, reporting their Explained Variance Ratio (EVR) to indicate the amount of variance captured by the visualization. Data points (representing individual audio samples) are color-coded: points with anomaly scores below a defined threshold are marked as 'normal', while those exceeding it are marked as 'anomaly'.

\subsection{Latent Representation Inspection (Proposed Models)}\label{sec:latent_representation_inspection}

To better understand the representations learned by our proposed \textit{AE} and \textit{AEwRES} models within the Deep SVDD framework, we perform two additional analyses:

\begin{itemize} 
  \item Latent Dimension Distributions: We plot density histograms of the activation values for each individual dimension in the learned latent space. This reveals the dataset's distribution, scale, and centering of values within each learned feature. 
  \item Latent Activation Heatmap: A heatmap visualizes the raw activation values across multiple samples (rows) and all latent dimensions (columns). This allows for inspecting activation patterns and consistency across samples and identifying potentially inactive or highly active latent dimensions. 
\end{itemize} 

These visualizations provide insights into the internal structure and characteristics of the features learned by models.

\subsection{Anomaly Score Distribution}\label{sec:anomaly_score_distribution}

We analyze the distribution of anomaly scores generated by each model (Deep SVDD variants and baselines). Anomaly scores for Deep SVDD models are the Euclidean distances from the hypersphere center ($c$), while baselines use their respective metrics (e.g., IF scores, PCA reconstruction errors). We plot overlaid histograms of these scores for the training and validation sets, using a logarithmic scale for the frequency axis to better visualize the distribution tails. The anomaly threshold value is calculated as the 95th percentile ($q=0.95$) of the anomaly scores obtained from the \textbf{training data} for each specific model. This threshold is depicted as a prominent vertical dashed line on the histograms. Its purpose is twofold: (1) it provides the reference for color-coding points in the PCA visualization, and (2) it visually demarcates the expected range for 'normal' data (concentrated below the threshold) from potentially 'anomalous' samples (scoring above it). This allows a comparison of the model's sensitivity and the separation achieved between training and validation score distributions.

Dimensionality reduction (PCA) and baseline modeling (IF, PCA reconstruction error) are performed using \texttt{scikit-learn}. Appendix \ref{app:pca_reconstruction_error} presents the implementation details for the PCA reconstruction error baseline method. All visualizations, including scatter plots, histograms, and heatmaps, are generated using \texttt{Matplotlib} and \texttt{Seaborn} Python libraries.

\section{Results}\label{sec:results}

\begin{figure}[t]
  \centering
  \includegraphics[width=\textwidth]{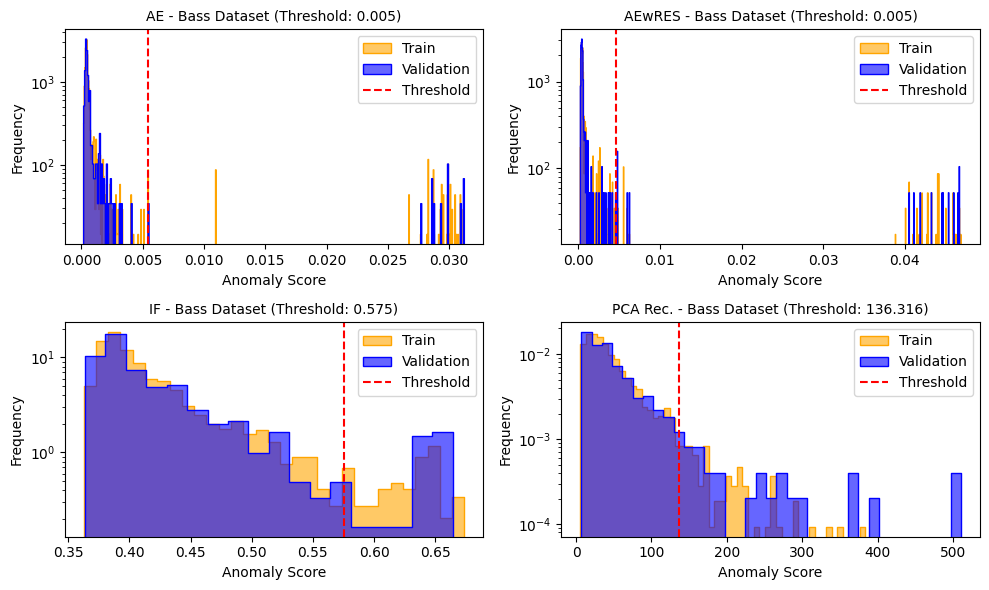}
  \caption{The figure displays the anomaly score distribution histograms for models applied to the bass dataset. The top row shows the proposed models (\textit{AE} and \textit{AEwRES}) and the bottom row displays baseline models (Section \ref{sec:baseline_models}). Each plot depicts the frequency distribution of anomaly scores for both training data (yellow) and validation data (blue), with red dashed lines indicating the anomaly threshold set at the 95th percentile of the training scores. The thresholds are specifically labeled for each model: 0.005 for both \textit{AE} and \textit{AEwRES}, 0.575 for IF, and 136.316 for PCA Reconstruction error.  \label{fig:benchmark_scores_bass}}
\end{figure}

\begin{figure}[t]
  \centering
  \includegraphics[width=\textwidth]{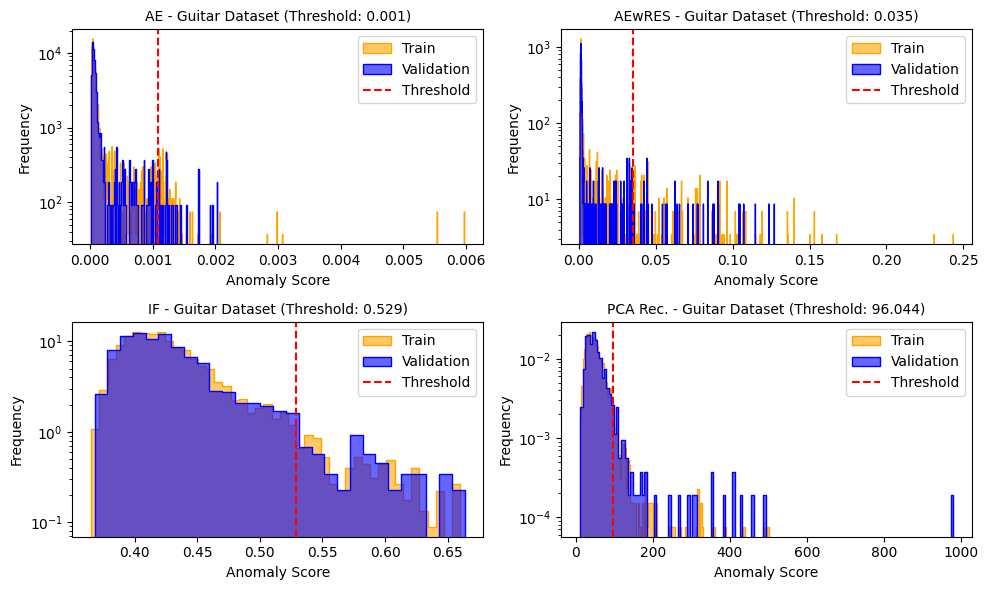}
  \caption{The figure displays anomaly score distribution histograms for models applied to the Guitar dataset. The top row shows the proposed models (AE and AEwRES) and the bottom row displays baseline models (Section \ref{sec:baseline_models}). Each plot depicts the frequency distribution of anomaly scores for both training data (yellow) and validation data (blue), with red dashed lines indicating the anomaly threshold set at the 95th percentile of the training scores. The thresholds are specifically labeled for each model: 0.001 for AE and 0.035 for AEwRES, 0.529 for IF, and 96.044 for PCA Reconstruction. \label{fig:benchmark_scores_guitar}}
\end{figure}

This section presents the findings of our evaluation, comparing the performance of the selected Deep SVDD variants (\textit{AE} and \textit{AEwRES}) against baseline methods (IF and PCA reconstruction error - Section \ref{sec:baseline_models}) on the bass and guitar datasets (Section \ref{sec:dataset}). Our selection of \textit{AE} and \textit{AEwRES}, utilizing the HTS-AT encoder with FFM, was informed by preliminary experiments (Appendix \ref{app:preliminary_experiment}), which confirmed the benefits of FFM and the selected architecture (Figure \ref{fig:aewres_architecture}). For the subsequent analysis, potential anomalies are defined as samples whose anomaly score exceeds the 95th percentile (q=0.95) threshold derived from the training set scores. Visualizations supporting this section are found in Figures \ref{fig:benchmark_scores_bass}, \ref{fig:benchmark_scores_guitar}, \ref{fig:anomaly_scores_distribution}, and Appendix \ref{app:main_experiments_results}. The implementation and supplementary materials will be provided in the accompaniment repository\footnote{\url{https://github.com/dadmaan/music-anomalizer.git}}.

\subsection{Performance on Bass Dataset}

On the bass dataset, both Deep SVDD models demonstrated effective separation capabilities, operating at lower anomaly score ranges than the baselines. The \textit{AE} model yielded a concentrated score distribution (mostly below 0.004, q=0.95, and threshold $\approx$ 0.005), creating a sharp distinction between normal samples and potential anomalies. The \textit{AEwRES} model showed a similarly compact distribution and threshold ($\approx$ 0.005), also achieving clear separation. In contrast, the IF operated around 0.45 (threshold $\approx$ 0.575), and PCA reconstruction error produced widely varying scores (10-500 range with threshold $\approx$ 136.3), indicating less precise definitions of normalcy.

Comparing the Deep SVDD latent representations (Appendix \ref{app:main_experiments_results}), \textit{AEwRES} produced a particularly tight clustering of normal data, evidenced by sharp KDE peaks (density $\approx$ 120) and clear PCA separation. The high explained variance by PC1 (89.6\%) and the structured heatmap patterns further underscore this. This suggests \textit{AEwRES} captures an enhanced representation of 'normal' bass patterns, facilitating the effective detection of deviations. The standard \textit{AE} model, while adequate, resulted in a broader KDE distribution (peak density $\approx$ 60) and less distinct PCA clustering, with lower PC1 variance (74.7\%). Its heatmap showed less contrast compared to \textit{AEwRES}. This implies a slightly less distinctive latent representation compared to \textit{AEwRES} for this dataset.

\begin{figure}[t]
  \centering
  \includegraphics[width=\textwidth]{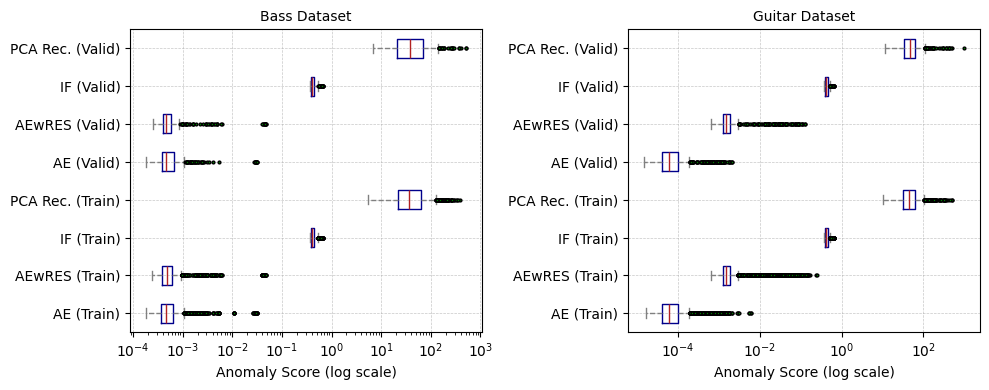}
  \caption{The figure displays box plots comparing anomaly score distributions across models (\textit{AE}, \textit{AEwRES}, baseline models (Section \ref{sec:baseline_models})) on Bass Dataset (left) and Guitar Dataset (right). Each panel shows eight box plots arranged vertically, with separate plots for training and validation data for each model. The anomaly scores are presented on a logarithmic scale. The plots reveal the statistical distribution of scores including medians, quartiles, and outliers for each model-dataset combination. \label{fig:anomaly_scores_distribution}}
\end{figure}

The box plot comparison (Figure \ref{fig:anomaly_scores_distribution}) visually confirms these findings: both \textit{AE} and \textit{AEwRES} assign considerably lower scores to normal data than baselines. \textit{AEwRES} exhibits a compact distribution, closely followed by \textit{AE}, reinforcing its ability to tightly model the normal manifold and separate outliers.

\subsection{Performance on Guitar Dataset}

Evaluation of the more complex and diverse guitar dataset revealed nuances in model performance. Again, the \textit{AE} model presented concentrated scores (below 0.001 with threshold $\approx$ 0.001). Conversely, \textit{AEwRES} appeared to accommodate the guitar data's inherent diversity (as elaborated in Section \ref{sec:dataset}) better, reflected in the broader score range (0-0.2 with threshold $\approx$ 0.035). Despite this range, its latent space exhibited structural properties indicative of effective modeling. The KDE plot showed distinct peaks capturing varied patterns, and the PCA projection revealed a curved cluster separated from potential anomalies. 

Furthermore, \textit{AEwRES} captured more variance within the first two principal components (93.2\% total: PC1=70.1\%, PC2=23.1\%) compared to \textit{AE} (77.7\%). This demonstrates a more informative latent representation learned by \textit{AEwRES} for the guitar samples. Its heatmap confirmed this ability to represent diverse yet structured normal patterns. In contrast, the \textit{AE} model's latent space for guitar data showed a sharply peaked, potentially over-compressed KDE (peak density $\approx$ 160) and less distinct separation in the PCA projection. This suggests \textit{AE}'s representation might over-simplify the guitar patterns, potentially hindering its ability to discern subtle anomalies. Its heatmap displayed more extreme activations, possibly overlooking nuanced variations, which are captured by \textit{AEwRES}. 

Compared to the baselines, both Deep SVDD models excelled. Unlike the baseline models, which struggled with the guitar data's diversity, leading to overlap between 'normal' and potential 'anomaly' representations (Appendix \ref{app:main_experiments_results}), both \textit{AE} and particularly \textit{AEwRES} maintained more precise separation. The structured latent space and effective thresholding of \textit{AEwRES} demonstrated more suitability and effectiveness for handling the guitar samples than both \textit{AE} and the baselines.

\subsection{Compartive Performance Summary}

As demonstrated across both datasets, the Deep SVDD approaches (\textit{AE} and \textit{AEwRES}) outperform the IF and PCA reconstruction error baselines. The baselines consistently exhibit higher anomaly scores for normal data and less distinct separation at their respective thresholds (Figure \ref{fig:anomaly_scores_distribution}). 

Between the two Deep SVDD variants, while the standard \textit{AE} model achieves high compactness, particularly on the simpler bass dataset, its tendency towards over-compression on the diverse guitar dataset suggests potential limitations. \textit{AEwRES} demonstrates enhanced feature separability capacity and better generalization. Its ability to model more intricate patterns while maintaining clear separation, particularly evident on the guitar dataset (Appendix \ref{app:main_experiments_results}), indicates its strength in learning distinctive latent representations. Therefore, despite \textit{AE} offering marginal compactness advantages on simpler data, \textit{AEwRES} appears to be the more promising architecture for loop detection in varied musical contexts due to its enhanced separation capabilities and ability to handle pattern diversity.

\section{Discussion}\label{sec:discussion}

Our investigation into automated loop detection confirms the effectiveness of the Deep SVDD framework, especially when coupled with our proposed \textit{AEwRES} encoder that integrates FFM and a HTS-AT. To rigorously evaluate its performance, we benchmarked this configuration against established anomaly detection techniques: IF and PCA-based reconstruction error (Section \ref{sec:baseline_models}). The rationale for including IF and PCA reconstruction error was to assess the added value of our deep, non-linear feature learning over simpler techniques for this specific task. Furthermore, to isolate the contributions of the residual connections within our encoder, we compared the \textit{AEwRES} model against a standard Autoencoder (\textit{AE}) variant operating within the same Deep SVDD paradigm.

Across both bass and guitar loop datasets, our proposed \textit{AEwRES} configuration demonstrated enhanced separative capacity compared to both the baselines (IF, PCA reconstruction error) and the \textit{AE} variant. This suggests that while general anomaly detectors provide a performance benchmark, the tailored feature representations learned by our architecture are effective for the nuances of loop detection. Specifically, the integration of FFM with HTS-AT successfully addresses the fixed-length input limitation inherent in previous studies (Section \ref{sec:related_works}). The success of FFM, as demonstrated in Appendix \ref{app:preliminary_experiment}, highlights its efficacy in aggregating temporal information, which is then transformed into meaningful representation by HTS-AT, facilitating the capture of both local and global temporal dependencies.

Comparative evaluations in Section \ref{sec:results} aimed to identify the most suitable architecture for loop detection. The observed performance variations between the bass and guitar datasets confirm the system's adaptability to individual user collections and specific stylistic nuances, supporting our initial hypothesis. On the bass dataset, both \textit{AE} and \textit{AEwRES} models learned compact representations of normal patterns. However, the \textit{AEwRES} model produced a slightly more structured latent space, suggesting an enhanced ability to capture discriminative features even with simpler data. This difference became more pronounced with the guitar dataset. The \textit{AE} model exhibited a tendency towards over-compression, potentially collapsing distinct normal patterns into an overly simplified representation, as evidenced by the lower explained variance in its latent space (77.7\% for the first two principal components) compared to the \textit{AEwRES} model (93.2\%).

Indeed, \textit{AEwRES} demonstrated more distinguished adaptability by modeling the guitar data's inherent variability through a more structured, albeit broader, latent representation, as evidenced by PCA visualizations and explained variance metrics. The architecture's effectiveness in creating a structured latent space allowed the Deep SVDD to define a meaningful hypersphere of normality, particularly for the complex guitar dataset. This suggests that architectures like \textit{AEwRES}, incorporating elements such as residual connections, are better equipped to handle the complexity and diversity commonly found in real-world musical audio.

Contextualizing these results, our application of Deep SVDD directly to audio features encoded by HTS-AT supports our hypothesis that such models can form the basis for loop detection and analysis, particularly when designed to capture nuanced temporal and hierarchical information. Compared to active learning systems requiring user feedback \citep{jakubik_searching_2022}, our unsupervised approach offers an automated alternative, though it lacks the user-guided refinement characteristic of such systems. The effectiveness of the learned representations aligns with findings emphasizing the practicality of pre-trained foundation models for audio tasks \citep{kong_panns_2020, wu_large-scale_2024}, and the \textit{AEwRES} architecture's performance echoes benefits seen from residual connections \citep{he_deep_2016} in other domains.

\subsection{Possible Applications}\label{sec:applications}

The advancements presented in this paper contribute to our broader research (Section \ref{sec:introduction}), which posits that MGS can evolve into genuinely adaptive and interactive creative partners capable of supporting music creation. By empowering the framework proposed in \citep{dadman2024Crafting} to effectively analyze variable-length audio loops (segments) and discern normative musical patterns without supervision, we enhance its 'musical intelligence' and practical versatility. This focus on understanding loops as musical segments directly tests and refines our overarching hypothesis, articulated in our foundational work \citep{dadman2022Interactive} and developed through \citep{dadman2024Crafting, dadman2023multiAgent}, concerning the co-creative potential of AI.

Moreover, the proposed approach presents several applications in music creation and audio analysis workflows, primarily in automated quality control and creative discovery. For quality control, the system can flag loops with unexpected artifacts, such as clicks, phase issues, or unintended noise. This would streamline the tedious auditing process for creators managing numerous takes or sample packs. Its unsupervised nature eliminates the need for pre-labeled 'anomalies', though effectiveness remains contingent on training data representativeness. In creative contexts, the method can be used for creative discovery through anomaly-driven sound design. By identifying loops that diverge from a user's existing library, the system could surface sonically unusual samples for experimental compositions, acting as a collaborator in sound exploration. For instance, a creator working on a glitch-hop track might use high anomaly scores to pinpoint loops with irregular rhythmic fractures or unexpected harmonic textures. This aligns with trends in "serendipitous retrieval" systems \citep{schedl2012Model, toms2001Serendipitous}.

Beyond these applications, the approach also holds potential for Digital Audio Workstations (DAWs) integration. Real-time loop (anomaly) detection could alert artists to problematic takes (e.g., a bass loop with inconsistent timing or a guitar loop with accidental string buzz) before they enter the production pipeline. A more speculative application involves style transfer and genre adaptation. By training the model on a corpus of loops from a specific genre (e.g., jazz guitar), users could detect deviations that signal cross-genre potential. Lastly, the method's hybrid approach—combining deep audio representations with unsupervised learning—circumvents the need for large labeled datasets, making it viable for independent creators with limited resources.

\section{Limitations and Future Work}\label{sec:limitations_future_work}

Despite the promising results and applications, several limitations must be acknowledged. Firstly, our evaluation relies on unsupervised anomaly detection, meaning we lack ground-truth labels for definitive anomalous loops. The 95th percentile threshold provides a heuristic separation point, but its optimality is data-dependent, and the interpretation of 'anomaly' remains relative to the training distribution. A truly 'anomalous' sound might be musically desirable in some contexts and undesirable noise in others. This leaves users to manually interpret whether an 'anomaly' represents a desirable stylistic blend or an irrelevant outlier. Secondly, our study was confined to bass and guitar loops from specific dataset. While these represent common instrumentation, the model's generalizability to other instruments (e.g., drums, vocals, synthesizers), genres, or audio types (e.g., full tracks, sound effects) remains unexplored. The dataset used, while containing variation, may not fully capture the breadth of real-world audio encountered by practitioners.

Furthermore, the technical limitations include the inherent assumptions of Deep SVDD, which aims to find a minimal hypersphere enclosing normal data in latent space. Complex, multi-modal distributions of normal data (as hinted at in the guitar samples in Section \ref{sec:results}) might not be perfectly captured by a single hypersphere. However, the flexibility of the deep encoder mitigates this to some extent compared to classical SVDD. While powerful, the choice of HTS-AT as the encoder also brings considerable computational complexity compared to simpler CNNs, which could be a barrier for resource-constrained applications. Finally, potential biases in the dataset composition (e.g., the prevalence of certain playing styles or recording qualities) could influence the learned representation of 'normalcy' and subsequent loop detection performance.

Building upon these findings and limitations, future research could proceed in several directions. Expanding the evaluation to a broader range of instruments, genres, and audio formats is essential for assessing the true generalizability of the \textit{AEwRES} - Deep SVDD approach. Incorporating datasets with at least partially annotated anomalies or employing semi-supervised learning techniques could allow for more rigorous quantitative evaluation and refinement of the anomaly detection threshold. Further architectural explorations could involve fine-tuning the HTS-AT model specifically for the loop detection task or investigating knowledge distillation techniques to create lighter-weight models suitable for real-time processing without sacrificing too much performance. Investigating the integration of this loop detection model with other MIR tools, such as automatic transcription or source separation systems, could lead to improved audio analysis workflows. For instance, identifying 'anomalous' segments could trigger a deeper analysis of only those parts of an audio file. Finally, developing user interfaces allows practitioners to interact with the loop detection system, perhaps by adjusting sensitivity or providing feedback on flagged segments, which would enhance its practical utility. This allows practitioners to dynamically recalibrate detection sensitivity based on creative intent, potentially addressing several interpretation challenges outlined above.

\section{Conclusion}\label{sec:conclusion}

In conclusion, this work demonstrated that Deep SVDD, coupled with an audio encoder like HTS-AT with FFM, offers a viable approach for loop detection. The \textit{AEwRES} variant, in particular, showed promise due to its ability to learn discriminative latent representations that accommodate the diversity inherent in complex musical data, compared to the selected baselines (IF and PCA reconstruction error) and the \textit{AE} variant. While limitations regarding evaluation methodology and dataset scope exist, the results indicated the potential of this approach to alleviate challenges in automated audio analysis. By providing a means to automatically identify deviations from normative patterns directly within the audio domain and handling variable-length inputs effectively, this research lays the groundwork for future investigations into more nuanced, interpretable, and widely applicable loop detection systems.

\bibliographystyle{unsrtnat}
\bibliography{references}

\newpage
\appendix

\section[\appendixname~\thesubsection]{PCA Reconstruction Error Algorithm}\label{app:pca_reconstruction_error}
\input{appendix_0_pca_reconstruction_error}

\newpage
\section[\appendixname~\thesubsection]{Preliminary Experiments}\label{app:preliminary_experiment}

\input{appendix_1_preliminary_experiment}

\newpage
\section[\appendixname~\thesubsection]{Main Experiment Supplementary Materials}\label{app:main_experiments_results}

\input{appendix_2_main_experiment}

\end{document}

%% file: appendix_0_pca_reconstruction_error.tex
Principal Component Analysis (PCA) based reconstruction error is a widely used technique for unsupervised anomaly detection \citep{chandola2009Anomaly, shyu2003Novel}. It operates under the assumption that the majority of the training data represents normal behavior, and that this normal data lies predominantly within a lower-dimensional subspace of the original feature space. Anomalies, conversely, are expected to deviate from this normal subspace. PCA is employed to identify this principal subspace from the training data. The anomaly score for any given data point is then calculated as the error incurred when attempting to reconstruct the point after projecting it onto this learned normal subspace. Points that deviate substantially from the normal patterns captured by the principal components will exhibit a high reconstruction error.

The complete implementation procedure is formally described in Algorithm \ref{alg:pca_reconstruction_error}.

\begin{algorithm}[ht]
    \caption{PCA Reconstruction Error for Anomaly Detection}
    \label{alg:pca_reconstruction_error}
    \begin{algorithmic}[1] 
    \Require Training data embeddings $\mathbf{X}_{\text{train}} \in \mathbb{R}^{n_{\text{train}} \times d}$
    \Require Evaluation data embeddings $\mathbf{X}_{\text{eval}} \in \mathbb{R}^{n_{\text{eval}} \times d}$
    \Require Number of components $n_{\text{components}}$ (integer, float, or None)
    \Require Variance threshold $\theta_{\text{var}}$ (default 0.95)
    \Require {\tt standardize} (boolean)

    \State Let $\mathbf{X}'_{\text{train}} = \mathbf{X}_{\text{train}}$ and $\mathbf{X}'_{\text{eval}} = \mathbf{X}_{\text{eval}}$
    \If{standardize is True}
        \State Compute mean $\boldsymbol{\mu}$ and standard deviation $\boldsymbol{\sigma}$ from $\mathbf{X}_{\text{train}}$
        \State Standardize: $\mathbf{X}'_{\text{train}} \gets (\mathbf{X}_{\text{train}} - \boldsymbol{\mu}) / \boldsymbol{\sigma}$
        \State Standardize: $\mathbf{X}'_{\text{eval}} \gets (\mathbf{X}_{\text{eval}} - \boldsymbol{\mu}) / \boldsymbol{\sigma}$
    \EndIf

    \If{$n_{\text{components}}$ is None}
        \State Fit PCA on $\mathbf{X}'_{\text{train}}$ (full rank)
        \State Compute cumulative explained variance ratios
        \State $k \gets \min\{j : \sum_{i=1}^j \text{variance}_i \geq \theta_{\text{var}}\}$
        \State $n_{\text{selected}} \gets k$
    \ElsIf{$n_{\text{components}}$ is float and $0 < n_{\text{components}} < 1$}
        \State $n_{\text{selected}} \gets n_{\text{components}}$  \Comment{as explained variance}
    \Else
        \State $n_{\text{selected}} \gets n_{\text{components}}$  \Comment{as integer}
    \EndIf

    \State Fit PCA model on $\mathbf{X}'_{\text{train}}$ with $n_{\text{selected}}$ components

    \State Initialize empty lists $\mathbf{e}_{\text{train}}$ and $\mathbf{e}_{\text{eval}}$

    \For{each $\mathbf{x}$ in $\mathbf{X}'_{\text{train}}$} \Comment{Calculate training errors}
        \State Project: $\mathbf{z} \gets \Call{PCA\_transform}{\mathbf{x}}$
        \State Reconstruct: $\hat{\mathbf{x}} \gets \Call{PCA\_inverse\_transform}{\mathbf{z}}$
        \State Compute error: $e \gets \sum_{i=1}^d (x_i - \hat{x}_i)^2$
        \State Append $e$ to $\mathbf{e}_{\text{train}}$
    \EndFor

    \For{each $\mathbf{x}$ in $\mathbf{X}'_{\text{eval}}$} \Comment{Calculate evaluation errors}
        \State Project: $\mathbf{z} \gets \Call{PCA\_transform}{\mathbf{x}}$
        \State Reconstruct: $\hat{\mathbf{x}} \gets \Call{PCA\_inverse\_transform}{\mathbf{z}}$
        \State Compute error: $e \gets \sum_{i=1}^d (x_i - \hat{x}_i)^2$
        \State Append $e$ to $\mathbf{e}_{\text{eval}}$
    \EndFor

    \State \Return array of training errors $\mathbf{e}_{\text{train}}$, array of evaluation errors $\mathbf{e}_{\text{eval}}$
    \end{algorithmic}
\end{algorithm}

%% file: appendix_1_preliminary_experiment.tex
This appendix details preliminary experiments conducted to validate key design choices and select promising model architectures for the main evaluation presented in Section \ref{sec:results}. To facilitate rapid iteration and efficient hyperparameter exploration, these initial tests were performed on a smaller, representative subset of the bass dataset (described in Section \ref{sec:dataset}), comprising 392 bass loops from the MusicRadar catalog \footnote{\url{https://www.musicradar.com/news/tech/sampleradar-392-free-bass-guitar-samples-537264}}. The primary objectives were to:

\begin{enumerate}
  \item Evaluate the effectiveness of the FFM described in our methodology (Section \ref{sec:methodology}).
  \item Compare different network architectures to select the most promising candidates for the subsequent evaluation in Section \ref{sec:results}.
\end{enumerate}

\begin{figure}[H]
  \centering
  \begin{subfigure}[b]{0.9\linewidth}
    \includegraphics[width=\linewidth]{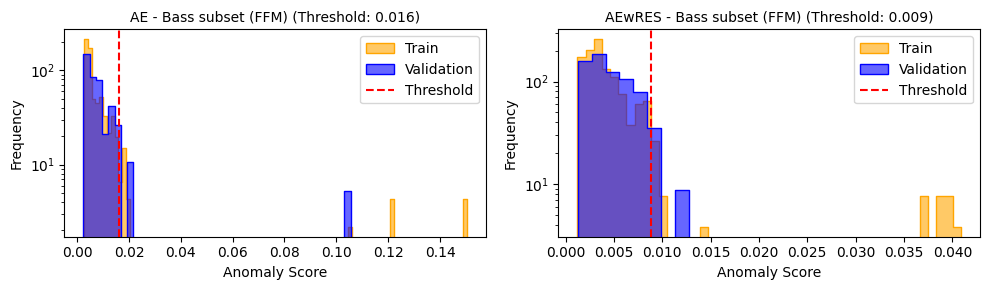}
    \caption{}
  \end{subfigure}
  \begin{subfigure}[b]{0.9\linewidth}
      \includegraphics[width=\linewidth]{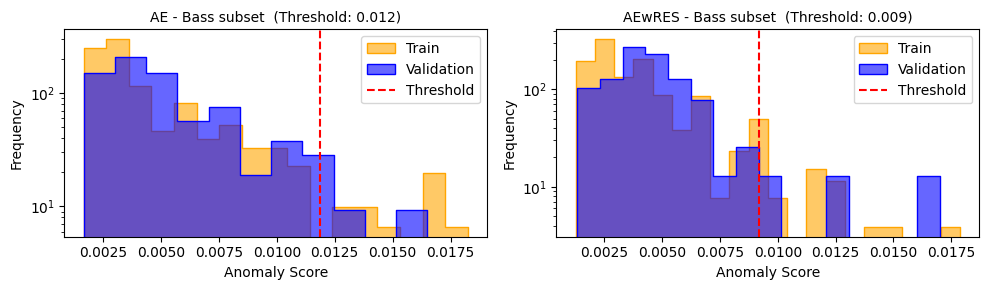}
      \caption{}
  \end{subfigure}
  \hfill
  \caption{The figure displays anomaly score distribution histograms for \textit{AE} and \textit{AEwRES} applied to the subset of bass dataset (392 samples). The top row (a) shows the scores obtained with FFM and the bottom row (b) displays the scores obtained witout FFM. Each plot depicts the frequency distribution of anomaly scores for both training data (yellow) and validation data (blue), with red dashed lines indicating the anomaly threshold set at the 95th percentile of the training scores.}
  \label{fig:pre_exp_anomaly_scores_392_nff_bass}
\end{figure}

\begin{figure}[H]
  \centering
  \begin{subfigure}[b]{0.85\linewidth}
    \includegraphics[width=\linewidth]{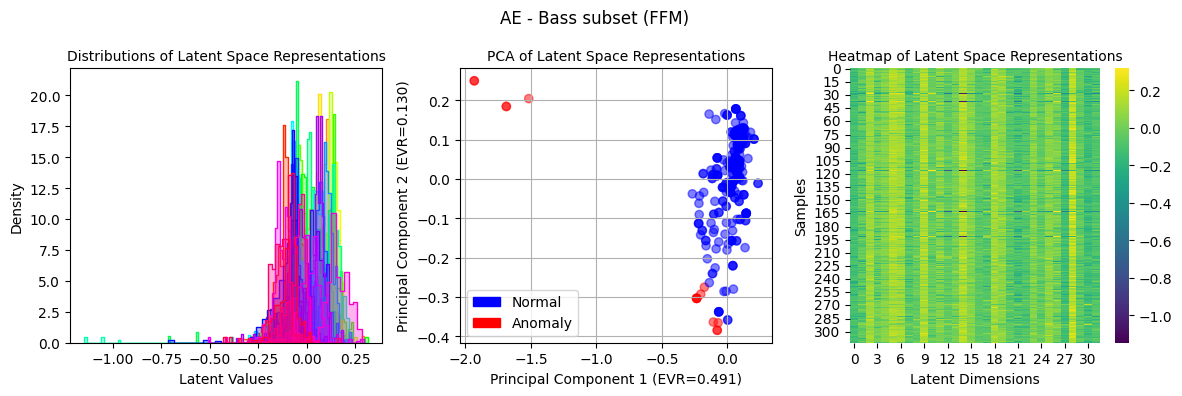}
    \caption{}
  \end{subfigure}
  \begin{subfigure}[b]{0.85\linewidth}
      \includegraphics[width=\linewidth]{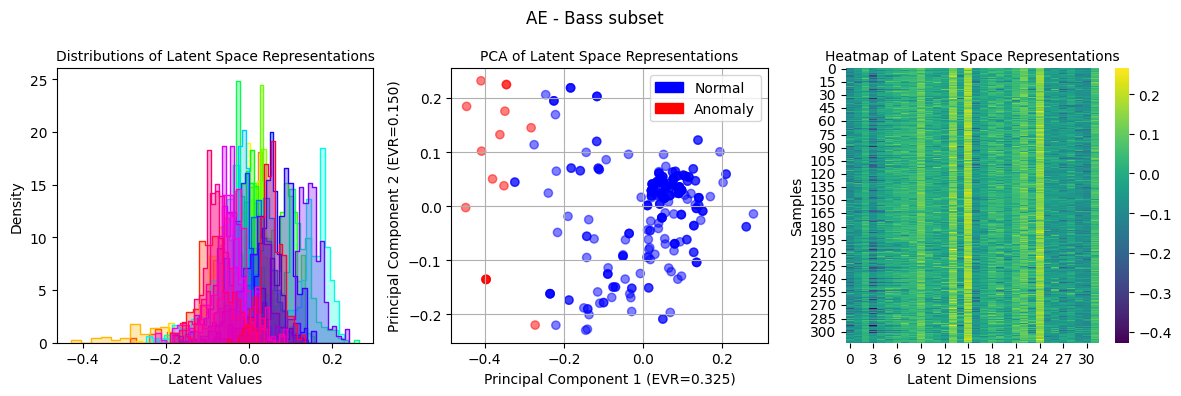}
      \caption{}
  \end{subfigure}
  \begin{subfigure}[b]{0.85\linewidth}
    \includegraphics[width=\linewidth]{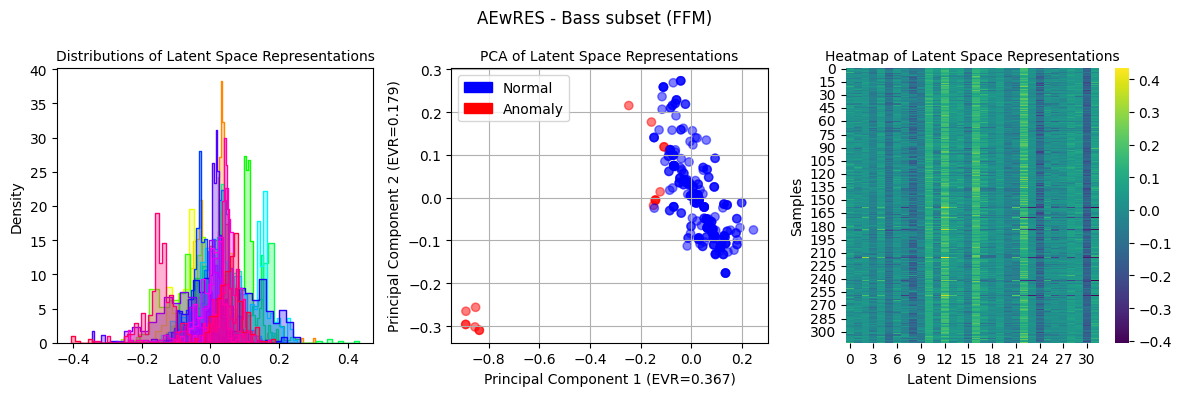}
    \caption{}
  \end{subfigure}
  \begin{subfigure}[b]{0.85\linewidth}
      \includegraphics[width=\linewidth]{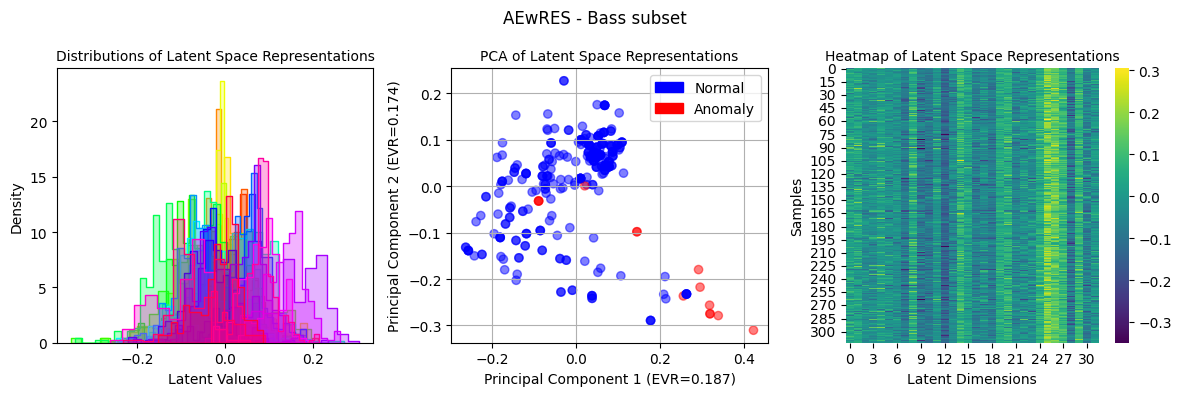}
      \caption{}
  \end{subfigure}
  \hfill
  \caption{The figure presents three visualizations of latent space representations from \textit{AE} and \textit{AEwRES} models applied to the subset of bass dataset (392 samples). The first and third rows (a, c) show the results obtained with FFM and the second and fourth rows (b, d) display the results obtained witout FFM. For each row, the left panel displays density distributions of latent values across multiple dimensions. The center panel shows a PCA scatter plot projecting the latent space onto two principal components, with blue points representing normal samples and red points indicating anomalies. The right panel features a heatmap of latent space representations across 32 dimensions (x-axis) for $N$ samples (y-axis), with color intensity reflecting latent values.}
  \label{fig:pre_exp_benchmark_392_bass}
\end{figure}

First, we assessed the impact of incorporating the FFM by comparing model variants (\textit{AE} and \textit{AEwRES}) trained with and without it. The results demonstrated the benefits of FFM. Models utilizing FFM converged faster and achieved improved representational quality. Specifically, FFM facilitated the models to capture underlying patterns, as evidenced by latent space distributions (KDE plots) and tighter clustering of normal samples in PCA projections (Figure \ref{fig:pre_exp_benchmark_392_bass}) and clearer separation in anomaly score distributions (Figure \ref{fig:pre_exp_anomaly_scores_392_nff_bass}) compared to models without it. Consequently, FFM was adopted for subsequent architecture comparisons and main experiments (Section \ref{sec:results}).

Following FFM validation, hyperparameter tuning was conducted for \textit{AE} and \textit{AEwRES} architectures to optimize performance. The tuned \textit{AE} exhibited a negatively skewed latent distribution ranging from -1.0 to 0.25, while \textit{AEwRES} showed a more balanced, symmetrical distribution centered around zero (-0.4 to 0.4). Feature representation heatmaps revealed higher contrast in \textit{AEwRES}, suggesting more distinct feature capture. Anomaly detection performance indicated successful separation of normal samples and potential anomalies in both architectures; however, \textit{AEwRES} achieved a lower anomaly threshold (q=0.95 $\approx$ 0.009) compared to \textit{AE} ($\approx$ 0.016), potentially indicating enhanced precision. During tuning, \textit{AEwRES} generally demonstrated faster convergence and more stable training dynamics. 

In summary, the optimized \textit{AEwRES} configuration showed advantages on this subset in terms of representation balance, feature distinctiveness, potential precision, and training efficiency. Nonetheless, given that \textit{AE} also performed competently after tuning and represents a different architectural approach, we selected both configurations for the further evaluations on the full datasets presented in Section \ref{sec:results}.

%% file: appendix_2_main_experiment.tex
\begin{figure}[h]
  \centering
  \begin{subfigure}[b]{0.85\linewidth}
    \includegraphics[width=\linewidth]{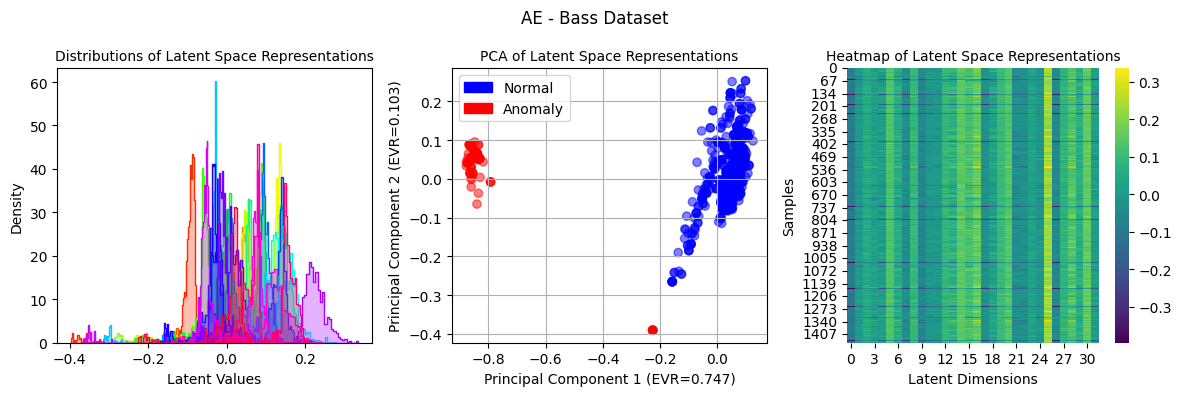}
    \caption{}
  \end{subfigure}
  \begin{subfigure}[b]{0.85\linewidth}
    \includegraphics[width=\linewidth]{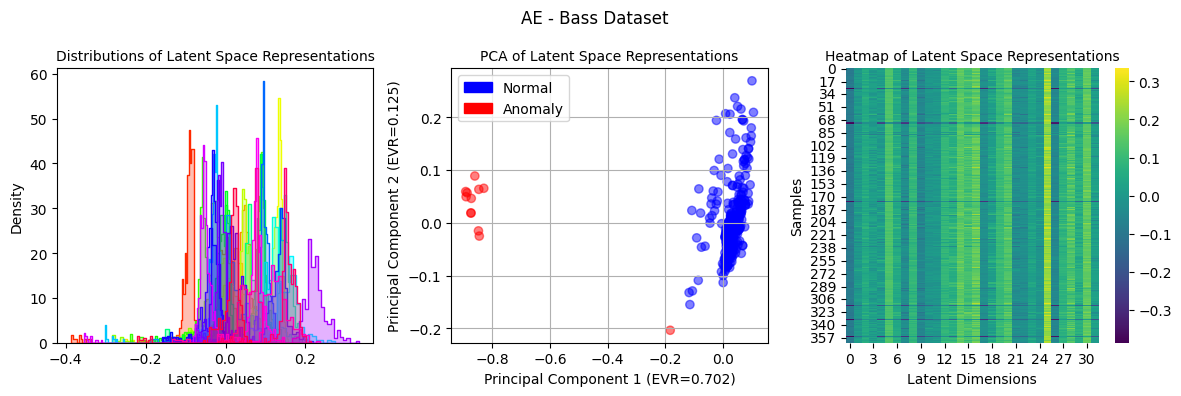}
    \caption{}
  \end{subfigure}
  \begin{subfigure}[b]{0.85\linewidth}
      \includegraphics[width=\linewidth]{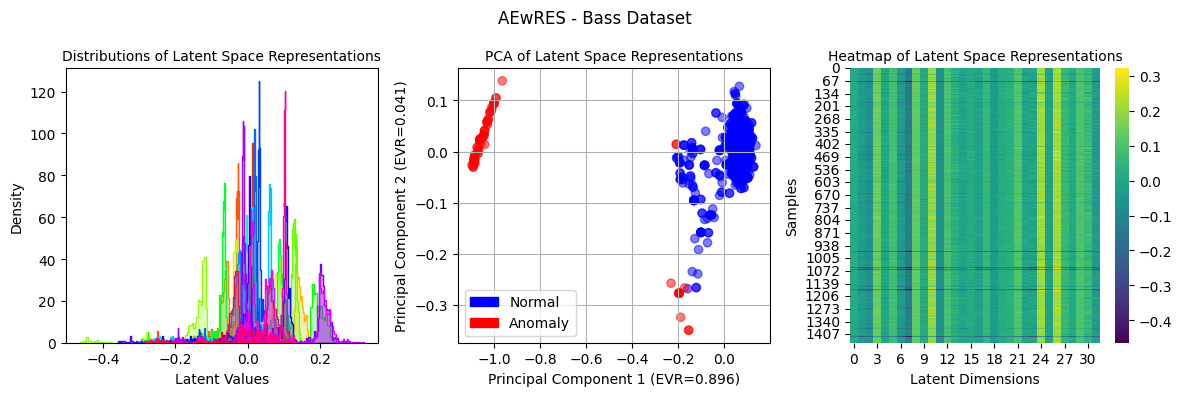}
      \caption{}
  \end{subfigure}
  \begin{subfigure}[b]{0.85\linewidth}
    \includegraphics[width=\linewidth]{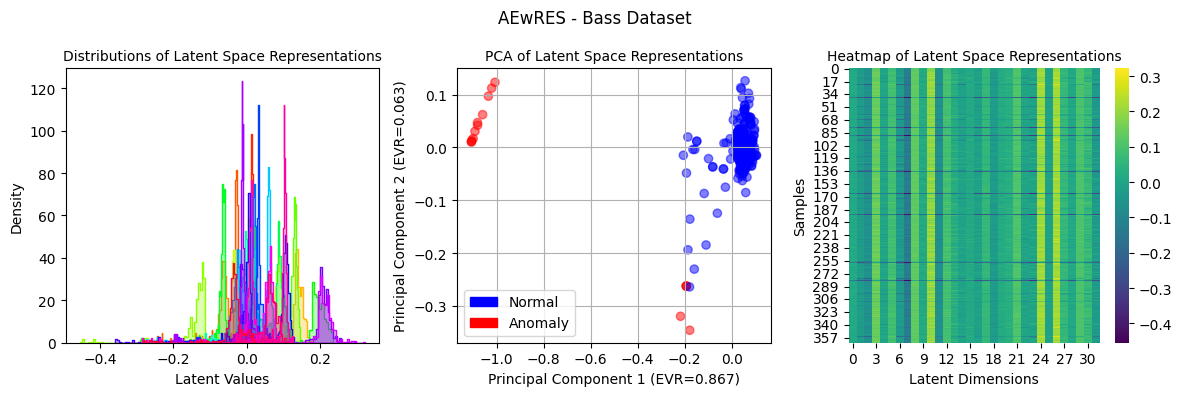}
    \caption{}
\end{subfigure}
  \hfill
  \caption{Comparative analysis of (a,b) \textit{AE} and (c,d) \textit{AEwRES} models using bass dataset, as described in Section \ref{sec:dataset}. The first two rows (a,b) show the results obtained by \textit{AE} model during (a) training and (b) validatio phases. The second two rows (c,d) displays the results obtained by \textit{AEwRES} model during (c) training and (d) validation phases. For each row, the left panel displays density distributions of latent values across multiple dimensions. The center panel shows a PCA scatter plot projecting the latent space onto two principal components, with blue points representing normal samples and red points indicating anomalies. The right panel features a heatmap of latent space representations across 32 dimensions (x-axis) for $N$ samples (y-axis), with color intensity reflecting latent values.}
  \label{fig:benchmark_bass}
\end{figure}
\begin{figure}[h]
  \centering
  \begin{subfigure}[b]{0.85\linewidth}
    \includegraphics[width=\linewidth]{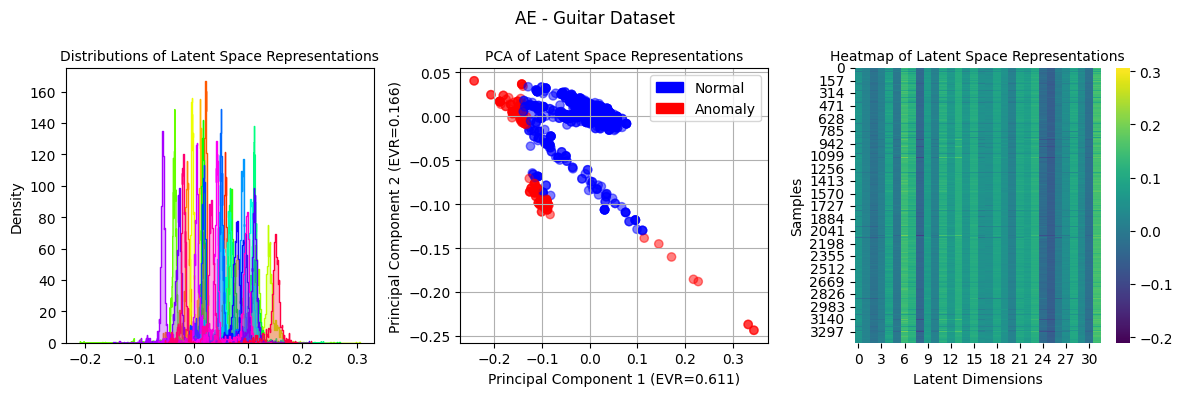}
    \caption{}
  \end{subfigure}
  \begin{subfigure}[b]{0.85\linewidth}
    \includegraphics[width=\linewidth]{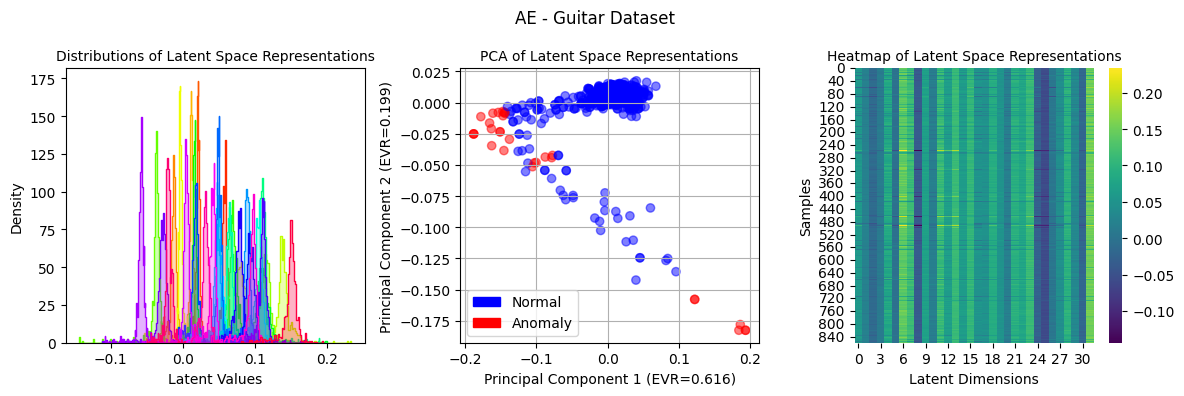}
    \caption{}
  \end{subfigure}
  \begin{subfigure}[b]{0.85\linewidth}
      \includegraphics[width=\linewidth]{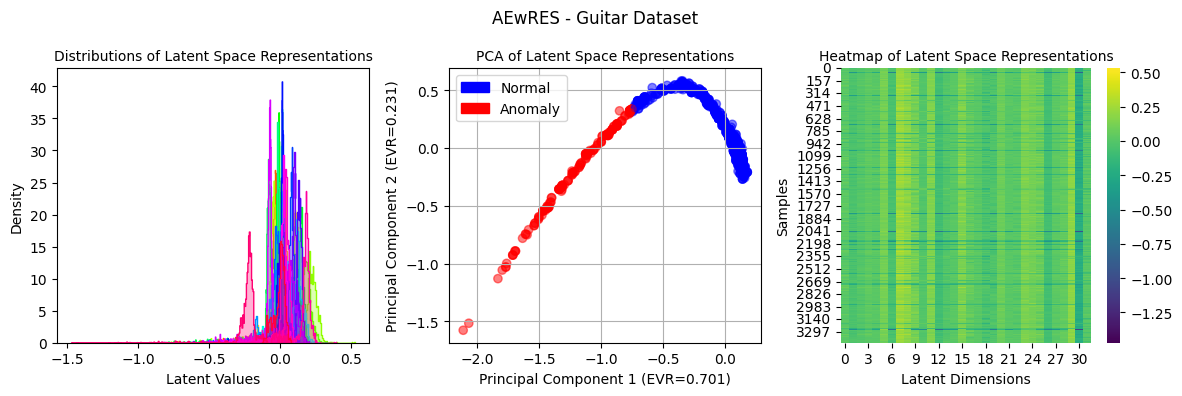}
      \caption{}
  \end{subfigure}
  \begin{subfigure}[b]{0.85\linewidth}
    \includegraphics[width=\linewidth]{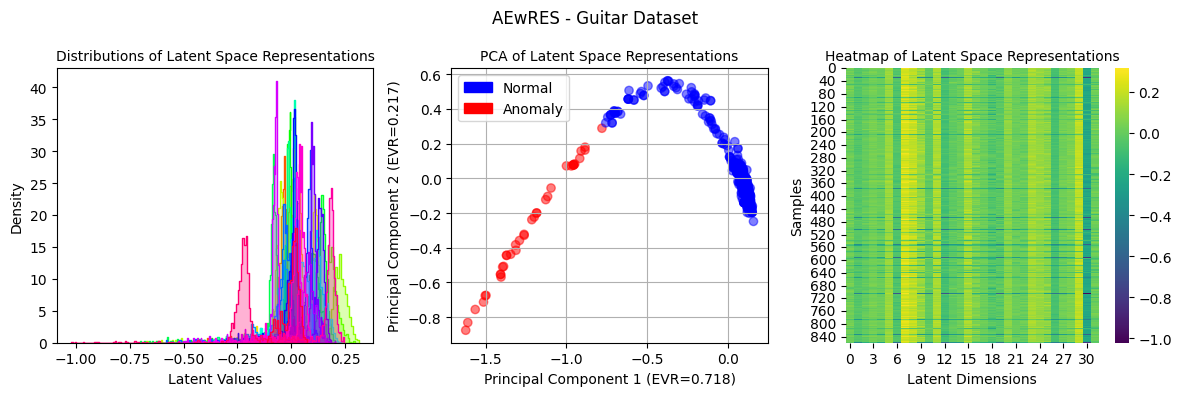}
    \caption{}
\end{subfigure}
  \hfill
  \caption{Comparative analysis of (a,b) \textit{AE} and (c,d) \textit{AEwRES} models using guitar dataset, as described in Section \ref{sec:dataset}. The first two rows (a,b) show the results obtained by \textit{AE} model during (a) training and (b) validatio phases. The second two rows (c,d) displays the results obtained by \textit{AEwRES} model during (c) training and (d) validatio phases. For each row, the left panel displays density distributions of latent values across multiple dimensions. The center panel shows a PCA scatter plot projecting the latent space onto two principal components, with blue points representing normal samples and red points indicating anomalies. The right panel features a heatmap of latent space representations across 32 dimensions (x-axis) for $N$ samples (y-axis), with color intensity reflecting latent values.}
  \label{fig:benchmark_guitar}
\end{figure}

This appendix provides additional visualizations for the results presented in Section \ref{sec:results}. Figures \ref{fig:benchmark_bass} and \ref{fig:benchmark_guitar} presented in this appendix illustrate the performance of the models on the bass and guitar datasets, respectively. They provide a more detailed analysis of the results, discussed in Section \ref{sec:results}. The figures show the density distributions of latent values across multiple dimensions, PCA scatter plots, and heatmaps of latent space representations, for training and vlidation phases. These visualizations are useful for understanding the behavior of the models and how they capture the underlying patterns in the data.

%% file: main.bbl
\begin{thebibliography}{57}
\providecommand{\natexlab}[1]{#1}
\providecommand{\url}[1]{\texttt{#1}}
\expandafter\ifx\csname urlstyle\endcsname\relax
  \providecommand{\doi}[1]{doi: #1}\else
  \providecommand{\doi}{doi: \begingroup \urlstyle{rm}\Url}\fi

\bibitem[Collins et~al.(2014)Collins, Schedel, and Wilson]{collins_electronic_2014}
Nick Collins, Margaret Schedel, and Scott Wilson.
\newblock \emph{Electronic music}.
\newblock Cambridge introductions to music. University Press, Cambridge, 1. publ edition, 2014.
\newblock ISBN 978-1-107-64817-3.

\bibitem[Butler(2006)]{butler_unlocking_2006}
Mark~J. Butler.
\newblock \emph{Unlocking the groove: rhythm, meter, and musical design in electronic dance music}.
\newblock Profiles in popular music. Indiana University Press, Bloomington, 2006.
\newblock ISBN 978-0-253-34662-9 978-0-253-21804-9.
\newblock OCLC: ocm61162077.

\bibitem[Reese et~al.(2009)Reese, Gross, and Gross]{reese_audio_2009}
David~E. Reese, Lynne~S. Gross, and Brian Gross.
\newblock \emph{Audio production worktext: concepts, techniques, and equipment}.
\newblock Elsevier Focal Press, Burlington, MA, 6th ed edition, 2009.
\newblock ISBN 978-0-240-81098-0.
\newblock OCLC: 255903242.

\bibitem[Gibson(2005)]{gibson_smrt_2005}
Bill Gibson.
\newblock \emph{The {S}.{M}.{A}.{R}.{T}. guide to producing music with samples, loops, and {MIDI}}.
\newblock Thomson Course Technology PTR : Artistpro Pub., Boston, MA, 2005.
\newblock ISBN 978-1-59200-697-7.
\newblock OCLC: 61519237.

\bibitem[Streich and Ong(2008)]{streich2008music}
Sebastian Streich and Bee~Suan Ong.
\newblock A music loop explorer system.
\newblock In \emph{ICMC}, 2008.

\bibitem[Kitahara et~al.(2015)Kitahara, Iijima, Okada, Yamashita, and Tsuruoka]{kitahara2015loop}
Tetsuro Kitahara, Kosuke Iijima, Misaki Okada, Yuji Yamashita, and Ayaka Tsuruoka.
\newblock A loop sequencer that selects music loops based on the degree of excitement.
\newblock In \emph{Proceedings of the 12th Sound and Music Computing Conference (SMC 2015)}, pages 435--438, 2015.

\bibitem[Chen et~al.(2020)Chen, Smith, and Yang]{chen_neural_2020}
Bo-Yu Chen, Jordan B.~L. Smith, and Yi-Hsuan Yang.
\newblock Neural {Loop} {Combiner}: {Neural} {Network} {Models} for {Assessing} the {Compatibility} of {Loops}, 2020.
\newblock URL \url{https://arxiv.org/abs/2008.02011}.
\newblock Version Number: 2.

\bibitem[Deruty et~al.(2022)Deruty, Grachten, Lattner, Nistal, and Aouameur]{deruty2022Development}
Emmanuel Deruty, Maarten Grachten, Stefan Lattner, Javier Nistal, and Cyran Aouameur.
\newblock On the {{Development}} and {{Practice}} of {{AI Technology}} for {{Contemporary Popular Music Production}}.
\newblock 5\penalty0 (1):\penalty0 35--49, 2022.
\newblock ISSN 2514-3298.
\newblock \doi{10.5334/tismir.100}.
\newblock URL \url{https://transactions.ismir.net/articles/10.5334/tismir.100?utm_source=TrendMD&utm_medium=cpc&utm_campaign=Transactions_of_the_International_Society_for_Music_Information_Retrieval_TrendMD_0}.

\bibitem[Huang et~al.(2020)Huang, Koops, Newton-Rex, Dinculescu, and Cai]{huang2020AI}
Cheng-Zhi~Anna Huang, Hendrik~Vincent Koops, Ed~Newton-Rex, Monica Dinculescu, and Carrie~J. Cai.
\newblock {{AI Song Contest}}: {{Human-AI Co-Creation}} in {{Songwriting}}, 2020.
\newblock URL \url{http://arxiv.org/abs/2010.05388}.

\bibitem[Dadman et~al.(2022)Dadman, Bremdal, Bang, and Dalmo]{dadman2022Interactive}
Shayan Dadman, Bernt~Arild Bremdal, Børre Bang, and Rune Dalmo.
\newblock Toward {{Interactive Music Generation}}: {{A Position Paper}}.
\newblock 10:\penalty0 125679--125695, 2022.
\newblock ISSN 2169-3536.
\newblock \doi{10.1109/ACCESS.2022.3225689}.
\newblock URL \url{https://ieeexplore.ieee.org/document/9966445/}.

\bibitem[Dadman and Bremdal(2024)]{dadman2024Crafting}
Shayan Dadman and Bernt~Arild Bremdal.
\newblock Crafting {{Creative Melodies}}: {{A User-Centric Approach}} for {{Symbolic Music Generation}}.
\newblock 13\penalty0 (6):\penalty0 1116, 2024.
\newblock ISSN 2079-9292.
\newblock \doi{10.3390/electronics13061116}.
\newblock URL \url{https://www.mdpi.com/2079-9292/13/6/1116}.

\bibitem[Dadman and Bremdal(2023)]{dadman2023multiAgent}
Shayan Dadman and Bernt~Arild Bremdal.
\newblock Multi-agent {Reinforcement} {Learning} for {Structured} {Symbolic} {Music} {Generation}.
\newblock In Philippe Mathieu, Frank Dignum, Paulo Novais, and Fernando De~La~Prieta, editors, \emph{Advances in {Practical} {Applications} of {Agents}, {Multi}-{Agent} {Systems}, and {Cognitive} {Mimetics}. {The} {PAAMS} {Collection}}, volume 13955, pages 52--63. Springer Nature Switzerland, Cham, 2023.
\newblock ISBN 978-3-031-37615-3 978-3-031-37616-0.
\newblock \doi{10.1007/978-3-031-37616-0_5}.
\newblock URL \url{https://link.springer.com/10.1007/978-3-031-37616-0_5}.
\newblock Series Title: Lecture Notes in Computer Science.

\bibitem[Civit et~al.(2022)Civit, Civit-Masot, Cuadrado, and Escalona]{civit2022Systematic}
Miguel Civit, Javier Civit-Masot, Francisco Cuadrado, and Maria~J. Escalona.
\newblock A systematic review of artificial intelligence-based music generation: {{Scope}}, applications, and future trends.
\newblock 209:\penalty0 118190, 2022.
\newblock ISSN 09574174.
\newblock \doi{10.1016/j.eswa.2022.118190}.
\newblock URL \url{https://linkinghub.elsevier.com/retrieve/pii/S0957417422013537}.

\bibitem[Fu et~al.(2011)Fu, Lu, Ting, and Zhang]{fu2011Survey}
Zhouyu Fu, Guojun Lu, Kai~Ming Ting, and Dengsheng Zhang.
\newblock A {{Survey}} of {{Audio-Based Music Classification}} and {{Annotation}}.
\newblock 13\penalty0 (2):\penalty0 303--319, 2011.
\newblock ISSN 1520-9210, 1941-0077.
\newblock \doi{10.1109/TMM.2010.2098858}.
\newblock URL \url{http://ieeexplore.ieee.org/document/5664796/}.

\bibitem[Ras and Wieczorkowska(2010)]{ras2010Advances}
Zbigniew~W. Ras and Alicja~A. Wieczorkowska, editors.
\newblock \emph{Advances in {{Music Information Retrieval}}}, volume 274 of \emph{Studies in {{Computational Intelligence}}}.
\newblock Springer Berlin Heidelberg, 2010.
\newblock ISBN 978-3-642-11673-5 978-3-642-11674-2.
\newblock \doi{10.1007/978-3-642-11674-2}.
\newblock URL \url{https://link.springer.com/10.1007/978-3-642-11674-2}.

\bibitem[Kong et~al.(2020)Kong, Cao, Iqbal, Wang, Wang, and Plumbley]{kong_panns_2020}
Qiuqiang Kong, Yin Cao, Turab Iqbal, Yuxuan Wang, Wenwu Wang, and Mark~D. Plumbley.
\newblock {PANNs}: {Large}-{Scale} {Pretrained} {Audio} {Neural} {Networks} for {Audio} {Pattern} {Recognition}, August 2020.
\newblock URL \url{http://arxiv.org/abs/1912.10211}.
\newblock arXiv:1912.10211 [cs, eess].

\bibitem[Xie and Virtanen(2020)]{xie_zero-shot_2020}
Huang Xie and Tuomas Virtanen.
\newblock Zero-{Shot} {Audio} {Classification} via {Semantic} {Embeddings}, 2020.
\newblock URL \url{https://arxiv.org/abs/2011.12133}.
\newblock Version Number: 2.

\bibitem[Gemmeke et~al.(2017)Gemmeke, Ellis, Freedman, Jansen, Lawrence, Moore, Plakal, and Ritter]{gemmeke_audio_2017}
Jort~F. Gemmeke, Daniel P.~W. Ellis, Dylan Freedman, Aren Jansen, Wade Lawrence, R.~Channing Moore, Manoj Plakal, and Marvin Ritter.
\newblock Audio {Set}: {An} ontology and human-labeled dataset for audio events.
\newblock In \emph{2017 {IEEE} {International} {Conference} on {Acoustics}, {Speech} and {Signal} {Processing} ({ICASSP})}, pages 776--780, New Orleans, LA, March 2017. IEEE.
\newblock ISBN 978-1-5090-4117-6.
\newblock \doi{10.1109/ICASSP.2017.7952261}.
\newblock URL \url{http://ieeexplore.ieee.org/document/7952261/}.

\bibitem[Choi et~al.(2016)Choi, Fazekas, and Sandler]{choi_automatic_2016}
Keunwoo Choi, George Fazekas, and Mark Sandler.
\newblock Automatic tagging using deep convolutional neural networks, 2016.
\newblock URL \url{https://arxiv.org/abs/1606.00298}.
\newblock Version Number: 1.

\bibitem[{Dash, Sukanta Kumar} et~al.(2023){Dash, Sukanta Kumar}, {Solanki, S S}, and {Chakraborty, Soubhik}]{dashsukantakumar2023Comprehensive}
{Dash, Sukanta Kumar}, {Solanki, S S}, and {Chakraborty, Soubhik}.
\newblock A {{Comprehensive Review}} on {{Audio}} based {{Musical Instrument Recognition}}: {{Human-Machine Interaction Towards Industry}} 4.0.
\newblock 82\penalty0 (01), 2023.
\newblock ISSN 00224456, 09751084.
\newblock \doi{10.56042/jsir.v82i1.70251}.
\newblock URL \url{http://op.niscair.res.in/index.php/JSIR/article/view/70251}.

\bibitem[Wu et~al.(2024)Wu, Chen, Zhang, Hui, Nezhurina, Berg-Kirkpatrick, and Dubnov]{wu_large-scale_2024}
Yusong Wu, Ke~Chen, Tianyu Zhang, Yuchen Hui, Marianna Nezhurina, Taylor Berg-Kirkpatrick, and Shlomo Dubnov.
\newblock Large-scale {Contrastive} {Language}-{Audio} {Pretraining} with {Feature} {Fusion} and {Keyword}-to-{Caption} {Augmentation}, March 2024.
\newblock URL \url{http://arxiv.org/abs/2211.06687}.
\newblock arXiv:2211.06687 [cs, eess].

\bibitem[Lu et~al.(2004)Lu, Wang, and Zhang]{lu_repeating_2004}
Lie Lu, Muyuan Wang, and Hong-Jiang Zhang.
\newblock Repeating pattern discovery and structure analysis from acoustic music data.
\newblock In \emph{Proceedings of the 6th {ACM} {SIGMM} international workshop on {Multimedia} information retrieval}, pages 275--282, New York NY USA, October 2004. ACM.
\newblock ISBN 978-1-58113-940-2.
\newblock \doi{10.1145/1026711.1026756}.
\newblock URL \url{https://dl.acm.org/doi/10.1145/1026711.1026756}.

\bibitem[Ong et~al.(2006)]{ong2006structural}
Bee~Suan Ong et~al.
\newblock \emph{Structural analysis and segmentation of music signals}.
\newblock Citeseer, 2006.

\bibitem[Paulus et~al.(2010)Paulus, M{\"u}ller, and Klapuri]{paulus2010state}
Jouni Paulus, Meinard M{\"u}ller, and Anssi Klapuri.
\newblock State of the art report: Audio-based music structure analysis.
\newblock 2010.

\bibitem[Nieto et~al.(2020)Nieto, Mysore, Wang, Smith, Schlüter, Grill, and McFee]{nieto_audio-based_2020}
Oriol Nieto, Gautham~J. Mysore, Cheng-i Wang, Jordan B.~L. Smith, Jan Schlüter, Thomas Grill, and Brian McFee.
\newblock Audio-{Based} {Music} {Structure} {Analysis}: {Current} {Trends}, {Open} {Challenges}, and {Applications}.
\newblock \emph{Transactions of the International Society for Music Information Retrieval}, 3\penalty0 (1):\penalty0 246--263, December 2020.
\newblock ISSN 2514-3298.
\newblock \doi{10.5334/tismir.54}.
\newblock URL \url{http://transactions.ismir.net/articles/10.5334/tismir.54/}.

\bibitem[Han et~al.(2022)Han, Ihm, Lee, and Lim]{han_symbolic_2022}
Sangjun Han, Hyeongrae Ihm, Moontae Lee, and Woohyung Lim.
\newblock Symbolic {Music} {Loop} {Generation} with {Neural} {Discrete} {Representations}, October 2022.
\newblock URL \url{http://arxiv.org/abs/2208.05605}.
\newblock arXiv:2208.05605 [cs, eess].

\bibitem[Jakubik(2022)]{jakubik_searching_2022}
Jan Jakubik.
\newblock Searching {For} {Loops} {And} {Sound} {Samples} {With} {Feature} {Learning}.
\newblock pages 13--18, September 2022.
\newblock \doi{10.15439/2022F279}.
\newblock URL \url{https://annals-csis.org/Volume_31/drp/279.html}.

\bibitem[Chen et~al.(2022)Chen, Du, Zhu, Ma, Berg-Kirkpatrick, and Dubnov]{chen_hts-at_2022}
Ke~Chen, Xingjian Du, Bilei Zhu, Zejun Ma, Taylor Berg-Kirkpatrick, and Shlomo Dubnov.
\newblock {HTS}-{AT}: {A} {Hierarchical} {Token}-{Semantic} {Audio} {Transformer} for {Sound} {Classification} and {Detection}, 2022.
\newblock URL \url{https://arxiv.org/abs/2202.00874}.
\newblock Version Number: 1.

\bibitem[Ruff et~al.(2018)Ruff, Vandermeulen, Goernitz, Deecke, Siddiqui, Binder, Muller, and Kloft]{ruff_deep_2018}
Lukas Ruff, Robert Vandermeulen, Nico Goernitz, Lucas Deecke, Shoaib~Ahmed Siddiqui, Alexander Binder, Emmanuel Muller, and Marius Kloft.
\newblock Deep one-class classification.
\newblock \emph{PMLR}, pages 4393--4402, 2018.

\bibitem[Xu et~al.(2023)Xu, Wang, Bi, Liu, and Wang]{xu2023SemiSupervised}
Liang Xu, Lizhong Wang, Sijun Bi, Hanyue Liu, and Jing Wang.
\newblock Semi-{{Supervised Sound Event Detection}} with {{Pre-Trained Model}}.
\newblock In \emph{{{ICASSP}} 2023 - 2023 {{IEEE International Conference}} on {{Acoustics}}, {{Speech}} and {{Signal Processing}} ({{ICASSP}})}, pages 1--5. IEEE, 2023.
\newblock ISBN 978-1-7281-6327-7.
\newblock \doi{10.1109/ICASSP49357.2023.10095687}.
\newblock URL \url{https://ieeexplore.ieee.org/document/10095687/}.

\bibitem[He et~al.(2016)He, Zhang, Ren, and Sun]{he_deep_2016}
Kaiming He, Xiangyu Zhang, Shaoqing Ren, and Jian Sun.
\newblock Deep {Residual} {Learning} for {Image} {Recognition}.
\newblock In \emph{2016 {IEEE} {Conference} on {Computer} {Vision} and {Pattern} {Recognition} ({CVPR})}, pages 770--778, Las Vegas, NV, USA, June 2016. IEEE.
\newblock ISBN 978-1-4673-8851-1.
\newblock \doi{10.1109/CVPR.2016.90}.
\newblock URL \url{http://ieeexplore.ieee.org/document/7780459/}.

\bibitem[Howard et~al.()Howard, Zhu, Chen, Kalenichenko, Wang, Weyand, Andreetto, and Adam]{howard2017MobileNets}
Andrew~G. Howard, Menglong Zhu, Bo~Chen, Dmitry Kalenichenko, Weijun Wang, Tobias Weyand, Marco Andreetto, and Hartwig Adam.
\newblock {{MobileNets}}: {{Efficient Convolutional Neural Networks}} for {{Mobile Vision Applications}}.
\newblock URL \url{http://arxiv.org/abs/1704.04861}.

\bibitem[Shi and Mysore(2018)]{shi2018LoopMaker}
Zhengshan Shi and Gautham~J. Mysore.
\newblock {{LoopMaker}}: {{Automatic}} creation of music loops from pre-recorded music.
\newblock In \emph{Proceedings of the 2018 {{CHI}} Conference on Human Factors in Computing Systems}, Chi '18, pages 1--6. Association for Computing Machinery, 2018.
\newblock ISBN 978-1-4503-5620-6.
\newblock \doi{10.1145/3173574.3174028}.
\newblock URL \url{https://doi-org.mime.uit.no/10.1145/3173574.3174028}.

\bibitem[Ong and Streich(2008)]{ong2008Music}
Bee~Suan Ong and Sebastian Streich.
\newblock Music loop extraction from digital audio signals.
\newblock In \emph{2008 {{IEEE International Conference}} on {{Multimedia}} and {{Expo}}}, pages 681--684. IEEE, 2008.
\newblock ISBN 978-1-4244-2570-9.
\newblock \doi{10.1109/ICME.2008.4607526}.
\newblock URL \url{http://ieeexplore.ieee.org/document/4607526/}.

\bibitem[Davies et~al.(2014)Davies, Hamel, Yoshii, and Goto]{davies2014AutoMashUpper}
Matthew E.~P. Davies, Philippe Hamel, Kazuyoshi Yoshii, and Masataka Goto.
\newblock {{AutoMashUpper}}: {{Automatic Creation}} of {{Multi-Song Music Mashups}}.
\newblock 22\penalty0 (12):\penalty0 1726--1737, 2014.
\newblock ISSN 2329-9290, 2329-9304.
\newblock \doi{10.1109/TASLP.2014.2347135}.
\newblock URL \url{http://ieeexplore.ieee.org/document/6876193/}.

\bibitem[Gebhardt et~al.(2016)Gebhardt, Davies, and Seeber]{gebhardt2016Psychoacoustic}
Roman Gebhardt, Matthew Davies, and Bernhard Seeber.
\newblock Psychoacoustic {{Approaches}} for {{Harmonic Music Mixing}}.
\newblock 6\penalty0 (5):\penalty0 123, 2016.
\newblock ISSN 2076-3417.
\newblock \doi{10.3390/app6050123}.
\newblock URL \url{https://www.mdpi.com/2076-3417/6/5/123}.

\bibitem[Bernardes et~al.(2017)Bernardes, Davies, and Guedes]{bernardes2017Perceptuallymotivateda}
Gilberto Bernardes, Matthew Davies, and Carlos Guedes.
\newblock A perceptually-motivated harmonic compatibility method for music mixing.
\newblock 2017.

\bibitem[Smith and Goto(2018)]{smith2018Nonnegative}
Jordan B.~L. Smith and Masataka Goto.
\newblock Nonnegative {{Tensor Factorization}} for {{Source Separation}} of {{Loops}} in {{Audio}}.
\newblock In \emph{2018 {{IEEE International Conference}} on {{Acoustics}}, {{Speech}} and {{Signal Processing}} ({{ICASSP}})}, pages 171--175. IEEE, 2018.
\newblock ISBN 978-1-5386-4658-8.
\newblock \doi{10.1109/ICASSP.2018.8461876}.
\newblock URL \url{https://ieeexplore.ieee.org/document/8461876/}.

\bibitem[Defferrard et~al.(2016)Defferrard, Benzi, Vandergheynst, and Bresson]{defferrard_fma_2016}
Michaël Defferrard, Kirell Benzi, Pierre Vandergheynst, and Xavier Bresson.
\newblock {FMA}: {A} {Dataset} {For} {Music} {Analysis}, 2016.
\newblock URL \url{https://arxiv.org/abs/1612.01840}.
\newblock Version Number: 3.

\bibitem[Grill et~al.(2020)Grill, Strub, Altché, Tallec, Richemond, Buchatskaya, Doersch, Pires, Guo, Azar, Piot, Kavukcuoglu, Munos, and Valko]{grill_bootstrap_2020}
Jean-Bastien Grill, Florian Strub, Florent Altché, Corentin Tallec, Pierre~H. Richemond, Elena Buchatskaya, Carl Doersch, Bernardo~Avila Pires, Zhaohan~Daniel Guo, Mohammad~Gheshlaghi Azar, Bilal Piot, Koray Kavukcuoglu, Rémi Munos, and Michal Valko.
\newblock Bootstrap your own latent: {A} new approach to self-supervised {Learning}, 2020.
\newblock URL \url{https://arxiv.org/abs/2006.07733}.
\newblock Version Number: 3.

\bibitem[Dai et~al.(2020)Dai, Gieseke, Oehmcke, Wu, and Barnard]{dai_attentional_2020}
Yimian Dai, Fabian Gieseke, Stefan Oehmcke, Yiquan Wu, and Kobus Barnard.
\newblock Attentional {Feature} {Fusion}, 2020.
\newblock URL \url{https://arxiv.org/abs/2009.14082}.
\newblock Version Number: 2.

\bibitem[Xin et~al.(2024)Xin, Zhu, Cheng, Yang, and Zou]{xin2024Audiotext}
Yifei Xin, Zhihong Zhu, Xuxin Cheng, Xusheng Yang, and Yuexian Zou.
\newblock Audio-text {{Retrieval}} with {{Transformer-based Hierarchical Alignment}} and {{Disentangled Cross-modal Representation}}, 2024.
\newblock URL \url{http://arxiv.org/abs/2409.09256}.

\bibitem[Xin et~al.(2023{\natexlab{a}})Xin, Yang, and Zou]{xin2023Backgroundaware}
Yifei Xin, Dongchao Yang, and Yuexian Zou.
\newblock Background-aware {{Modeling}} for {{Weakly Supervised Sound Event Detection}}.
\newblock In \emph{{{INTERSPEECH}} 2023}, pages 1199--1203. ISCA, 2023{\natexlab{a}}.
\newblock \doi{10.21437/Interspeech.2023-330}.
\newblock URL \url{https://www.isca-archive.org/interspeech_2023/xin23_interspeech.html}.

\bibitem[Xin et~al.(2023{\natexlab{b}})Xin, Yang, Cui, Wang, and Zou]{xin2023Improving}
Yifei Xin, Dongchao Yang, Fan Cui, Yujun Wang, and Yuexian Zou.
\newblock Improving {{Weakly Supervised Sound Event Detection}} with {{Causal Intervention}}, 2023{\natexlab{b}}.
\newblock URL \url{http://arxiv.org/abs/2303.05678}.

\bibitem[Tax and Duin(2004)]{tax_support_2004}
David~M.J. Tax and Robert~P.W. Duin.
\newblock Support {Vector} {Data} {Description}.
\newblock \emph{Machine Learning}, 54\penalty0 (1):\penalty0 45--66, January 2004.
\newblock ISSN 0885-6125.
\newblock \doi{10.1023/B:MACH.0000008084.60811.49}.
\newblock URL \url{http://link.springer.com/10.1023/B:MACH.0000008084.60811.49}.

\bibitem[{MusicRadar}(n.d.)]{musicradar_sampleradar}
{MusicRadar}.
\newblock Free music samples: royalty-free loops, hits, and multis to download - sampleradar, n.d.
\newblock URL \url{https://www.musicradar.com/news/tech/free-music-samples-royalty-free-loops-hits-and-multis-to-download-sampleradar}.
\newblock Accessed: 2024-10-04.

\bibitem[Foroughmand and Peeters(2019)]{foroughmand2019deeprhythm}
Hadrien Foroughmand and Geoffroy Peeters.
\newblock Deep-{Rhythm} for {Global} {Tempo} {Estimation} in {Music}.
\newblock November 2019.
\newblock \doi{10.5281/ZENODO.3527890}.
\newblock URL \url{https://zenodo.org/record/3527890}.
\newblock Publisher: Zenodo.

\bibitem[Rasamoelina et~al.(2020)Rasamoelina, Adjailia, and Sincak]{rasamoelina_review_2020}
Andrinandrasana~David Rasamoelina, Fouzia Adjailia, and Peter Sincak.
\newblock A {Review} of {Activation} {Function} for {Artificial} {Neural} {Network}.
\newblock In \emph{2020 {IEEE} 18th {World} {Symposium} on {Applied} {Machine} {Intelligence} and {Informatics} ({SAMI})}, pages 281--286, Herlany, Slovakia, January 2020. IEEE.
\newblock ISBN 978-1-72813-149-8.
\newblock \doi{10.1109/SAMI48414.2020.9108717}.
\newblock URL \url{https://ieeexplore.ieee.org/document/9108717/}.

\bibitem[Ronneberger et~al.(2015)Ronneberger, Fischer, and Brox]{ronneberger2015UNet}
Olaf Ronneberger, Philipp Fischer, and Thomas Brox.
\newblock U-{{Net}}: {{Convolutional Networks}} for {{Biomedical Image Segmentation}}, 2015.
\newblock URL \url{http://arxiv.org/abs/1505.04597}.

\bibitem[Loshchilov and Hutter(2019)]{loshchilov2019Decoupled}
Ilya Loshchilov and Frank Hutter.
\newblock Decoupled {{Weight Decay Regularization}}, 2019.
\newblock URL \url{http://arxiv.org/abs/1711.05101}.

\bibitem[Loshchilov and Hutter(2017)]{loshchilov2017SGDR}
Ilya Loshchilov and Frank Hutter.
\newblock {{SGDR}}: {{Stochastic Gradient Descent}} with {{Warm Restarts}}, 2017.
\newblock URL \url{http://arxiv.org/abs/1608.03983}.

\bibitem[Liu et~al.()Liu, Ting, and Zhou]{liu2008isolation}
Fei~Tony Liu, Kai~Ming Ting, and Zhi-Hua Zhou.
\newblock Isolation {{Forest}}.
\newblock In \emph{2008 {{Eighth IEEE International Conference}} on {{Data Mining}}}, pages 413--422. IEEE.
\newblock ISBN 978-0-7695-3502-9.
\newblock \doi{10.1109/ICDM.2008.17}.
\newblock URL \url{http://ieeexplore.ieee.org/document/4781136/}.

\bibitem[Jolliffe()]{jolliffe2002principal}
Ian~T. Jolliffe.
\newblock \emph{Principal Component Analysis}.
\newblock Springer Series in Statistics. Springer, 2nd ed edition.
\newblock ISBN 978-0-387-95442-4.
\newblock \doi{10.1007/b98835}.
\newblock URL \url{http://link.springer.com/10.1007/b98835}.

\bibitem[Schedl et~al.(2012)Schedl, Hauger, and Schnitzer]{schedl2012Model}
Markus Schedl, David Hauger, and Dominik Schnitzer.
\newblock A model for serendipitous music retrieval.
\newblock In \emph{Proceedings of the 2nd {{Workshop}} on {{Context-awareness}} in {{Retrieval}} and {{Recommendation}}}, pages 10--13. ACM, 2012.
\newblock ISBN 978-1-4503-1192-2.
\newblock \doi{10.1145/2162102.2162105}.
\newblock URL \url{https://dl.acm.org/doi/10.1145/2162102.2162105}.

\bibitem[Toms(2001)]{toms2001Serendipitous}
Elaine Toms.
\newblock Serendipitous information retrieval.
\newblock 2001.

\bibitem[Chandola et~al.(2009)Chandola, Banerjee, and Kumar]{chandola2009Anomaly}
Varun Chandola, Arindam Banerjee, and Vipin Kumar.
\newblock Anomaly detection: {{A}} survey.
\newblock 41\penalty0 (3):\penalty0 1--58, 2009.
\newblock ISSN 0360-0300, 1557-7341.
\newblock \doi{10.1145/1541880.1541882}.
\newblock URL \url{https://dl.acm.org/doi/10.1145/1541880.1541882}.

\bibitem[Shyu et~al.(2003)Shyu, Chen, Sarinnapakorn, and Chang]{shyu2003Novel}
Mei-Ling Shyu, Shu-Ching Chen, Kanoksri Sarinnapakorn, and Liwu Chang.
\newblock A novel anomaly detection scheme based on principal component classifier.
\newblock Proceedings of {{International Conference}} on {{Data Mining}}, 2003.

\end{thebibliography}
